\documentclass[article,twocolumn,prb,showpacs,superscriptaddress]{revtex4-2}

\usepackage{graphicx}
\usepackage{amssymb}
\usepackage{bm}
\usepackage{float}
\usepackage{amsmath}
\usepackage{xfrac}
\usepackage{color}
\usepackage{titlesec}
\renewcommand{\figurename}{Figure} 
\makeatletter	        
\def\fnum@figure{\textbf{\figurename~\thefigure}}
\makeatother
\usepackage[normalem]{ulem}
\usepackage[utf8]{inputenc}
\usepackage[T1]{fontenc}
\usepackage[marginal]{footmisc}

\usepackage{latexsym}

\usepackage[colorlinks=True,
                 linkcolor=black,
	        citecolor=blue,
	        urlcolor=blue]{hyperref}

\makeatletter	        
 \def\section{%
  \@startsection{section}{1}{\z@}{0.8cm plus1ex minus.2ex}{0.2cm}%
  {%
   \small\sffamily\bfseries\selectfont
   \raggedright
   \parindent\z@
  }%
 }%
  \def\subsection{%
  \@startsection{subsection}{2}{\z@}{0.8cm plus1ex minus.2ex}{0.2cm}%
  {%
   \small\sffamily\bfseries
   \raggedright
   \parindent\z@
  }%
 }%
\makeatother

\newcommand{\comment}{\textcolor{black}}

\usepackage{color}

\newcommand{\MOE}{MOE Key Laboratory for Nonequilibrium Synthesis and Modulation of Condensed Matter, Shaanxi Province Key Laboratory of Advanced Materials and Mesoscopic Physics, School of Physics, Xi’an Jiaotong University, Xi’an,710049, China}
\newcommand{\ME}{State Key Laboratory for Manufacturing Systems Engineering, Xi'an Jiaotong University, Xi’an,710049, China}
\newcommand{\SE}{Key Laboratory of Quantum Materials and Devices of Ministry of Education, School of Physics, Southeast University, 211189, Nanjing, China}
\newcommand{\JAP}{Research Center for Electronic and Optical Materials, National Institute for Materials Science, 1-1 Namiki, Tsukuba 305-0044, Japan}
\newcommand{\JAPP}{Research Center for Materials Nanoarchitectonics, National Institute for Materials Science,  1-1 Namiki, Tsukuba 305-0044, Japan}
\newcommand{\FD}{State Key Laboratory of Surface Physics and Institute for Nanoelectronic Devices and Quantum Computing, Fudan University, Shanghai, 200433, China}
\newcommand{\FDD}{Zhangjiang Fudan International Innovation Center, Fudan University, Shanghai 201210, China}

\makeatletter
\g@addto@macro\bfseries{\boldmath}
\makeatother

\begin{document}
\title{Magnetoresistance oscillations in vertical junctions of 2D antiferromagnetic semiconductor CrPS$_4$}
\author{Pengyuan Shi}
\thanks{These authors contribute equally to this work.}
\author{Xiaoyu Wang}
\thanks{These authors contribute equally to this work.}
\author{Lihao Zhang}
\affiliation{\MOE}
\author{Wenqin Song}
\affiliation{\FD}
\author{Kunlin Yang}
\affiliation{\FD}
\author{Shuxi Wang}
\author{Ruisheng Zhang}
\affiliation{\MOE}
\author{Liangliang Zhang}
\affiliation{\ME}
\author{Takashi Taniguchi}
\affiliation{\JAPP}
\author{Kenji Watanabe}
\affiliation{\JAP}
\author{Sen Yang}
\author{Lei Zhang}
\affiliation{\MOE}
\author{Lei Wang}
\affiliation{\SE}
\author{Wu Shi}
\affiliation{\FD}
\affiliation{\FDD}
\author{Jie Pan}
\email{jiepan@xjtu.edu.cn}
\author{Zhe Wang}
\email{zhe.wang@xjtu.edu.cn}
\affiliation{\MOE}



\begin{abstract}
Magnetoresistance (MR) oscillations serve as a hallmark of intrinsic quantum behavior, traditionally observed only in conducting systems. Here we report the discovery of MR oscillations in an insulating system, the vertical junctions of CrPS$_4$ which is a two dimensional (2D) A-type antiferromagnetic semiconductor. Systematic investigations of MR peaks under varying conditions, including electrode materials, magnetic field direction, temperature, voltage bias and layer number, elucidate a correlation between MR oscillations and spin-canted states in CrPS$_4$. Experimental data and analysis point out the important role of the in-gap electronic states in generating MR oscillations,  and we proposed that spin selected interlayer hopping of localized \comment{defect} states may be responsible for it. Our findings not only illuminate the unusual electronic transport in CrPS$_4$ but also underscore the potential of van der Waals magnets for exploring interesting phenomena.
\end{abstract}

\maketitle

\section*{I.	Introduction}

Magnetoresistance (MR) is a phenomenon widely studied for its demonstration of intriguing physics and potential applications in industrial technologies. Magnetoresistance oscillation, characterized by non-monotonic changes multi-times in resistance with respect to an external magnetic field, often manifests the inherent quantum mechanical nature of the studied system. The well-known examples include Shubnikov–de Haas  oscillations where the electron density of states on the Fermi surface varies with magnetic field, and Aharonov-Bohm effect where the phase conducting electrons shift by an external magnetic field~\cite{Solid}. However, these phenomena usually present in conducting systems, and observations of MR oscillations in insulating systems are rare.

MR of insulating magnetic systems actually has been a focus of research over the past decades, forming the foundation of spintronics, which has had a significant impact on current information technologies~\cite{RMP2004,RMP2020,Review2020}. One classic model is the magnetic tunneling junction, consisting of two ferromagnetic metals separated by a thin insulator. Electron transport in such systems occurs via tunneling, where resistance depends on the alignment of electron magnetizations~\cite{RMP2004,Review2020,tsymbal2003spin}. Another notable model is the spin filter of magnetic semiconductor, characterized by its spin-split band structure thus leading to different tunneling barrier heights for different spins. After the pioneering works on Eu-based systems~\cite{MooderaPRL,EuO}, recent advancements in 2D magnetic semiconductors~\cite{2D2018, 2D2019, 2D2019ZX, 2D2019Marco, 2D2020,CrSBrNM}, particularly in systems like CrI$_3$~\cite{CrI3Xu,CrI3Pablo,CrI3NC, CrI3}, have demonstrated substantial progress in achieving large magnetoresistance. In both magnetic tunneling junctions and spin filters with magnetic semiconductor, MR oscillation is not expected as resistance typically correlates monotonically with the total magnetization of the system.

Here we report observations of MR oscillations in vertical junctions of 2D antiferromagnetic semiconductor CrPS$_4$. While the total magnetization of few layer CrPS$_4$ is monotonically changed by magnetic field, several well-defined resistance peaks appear in the MR measurements. The MR oscillation is robust against different directions of applied magnetic field, and smoothly vanished as the temperature goes above Néel temperature of CrPS$_4$. No MR oscillation is observed when the junction barrier is thinned down to monolayer, which is a ferromagnetic semiconductor. These findings demonstrate the MR oscillations are correlated to the spin-canted states in the system. We further discuss the possible transport mechanism in CrPS$_4$ and point out the in-gap states would play an important role.  

\begin{figure*}
\centering
\includegraphics[width =0.9\linewidth]{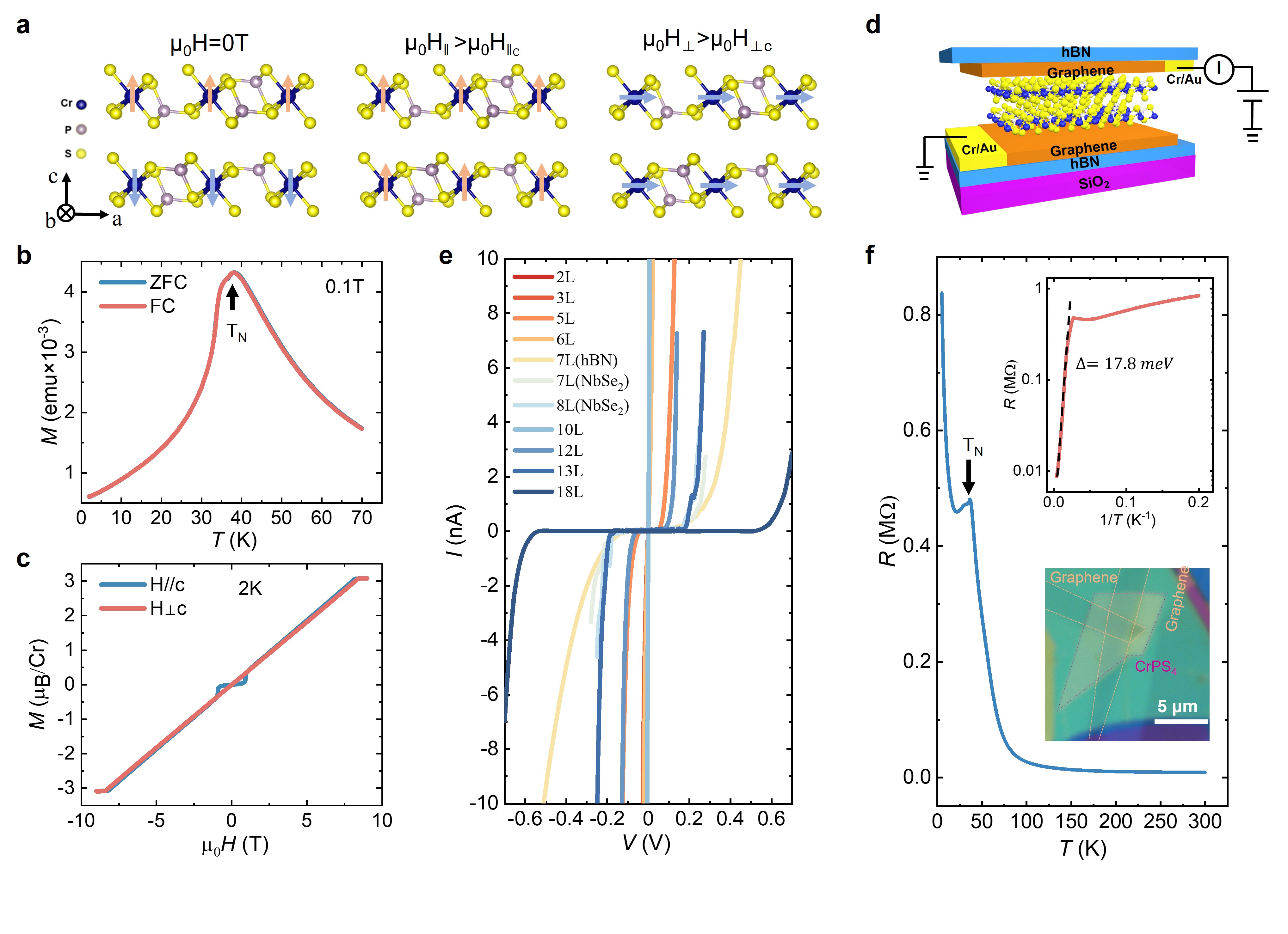}
\caption{\textbf{Basic characteristics of bulk CrPS$_4$ and vertical junctions of few-layer CrPS$_4$.} \textbf{a}. Crystal and magnetic structure of CrPS$_4$, depicting its A-type antiferromagnetic behavior below the Néel temperature with the easy axis along the $c$-axis. Above the spin-flip field, the magnetic moment aligns with the external magnetic field. \textbf{b}. Temperature dependence of magnetization under zero-field-cooling (ZFC) and field-cooling (FC) conditions, indicating a Néel temperature of around 38 K. \textbf{c}. Magnetic field dependence of magnetization with the magnetic field parallel and perpendicular to the $c$-axis. \textbf{d}. Schematic of vertical junctions, with graphene serving as electrodes and hBN used to encapsulate the entire device. \textbf{e}. $IV$ curves of different vertical junctions at 2 K, all devices exhibiting very large resistivity. Here 7L(hBN) denotes device with hBN inserted between graphene and CrPS$_4$, 7L(NbSe$_2$) and 8L(NbSe$_2$) represent devices with NbSe$_2$ as electrodes. All other devices use graphene as electrodes. \textbf{f}. Temperature dependence of resistance of a 10-layer device, with a kink clearly identified at the Néel temperature. The upper inset shows the Arrhenius plot, while the lower inset displays the optical image of the device.} 
\end{figure*}

\section*{II.	Experimental Results}

CrPS$_4$ is a van der Waals layered semiconductor with band gap of around 1.3 eV~\cite{CPSOld, ACSYeyu,CPSAPL,CPSPRB,CPSPRM,CPSNPJ}, the structure of which is depicted in Fig. 1a. Below the Néel temperature of 38 K (refer to Fig. 1b), Cr atoms within the ab plane of the crystal exhibit ferromagnetic coupling, while the interlayer coupling is antiferromagnetic~\cite{CPSPRB2021, ACSPark,AMYangjl2020}, being A-type antiferromagnet. Upon the application of a magnetic field along the $c$-axis, the spins undergo a spin-flop transition at approximately 0.7 Tesla, followed by a smooth alignment along the direction of the magnetic field until reaching the saturation field of approximately 8.25 Tesla. Conversely, when the magnetic field is applied perpendicular to the $c$-axis, no spin-flop transition occurs, and the magnetization increases smoothly with the magnetic field until reaching a saturation field of approximately 8.45 Tesla, as shown in Fig. 1c. These observations confirm the easy axis of CrPS$_4$ to be along the $c$-axis, consistent with previous bulk measurements~\cite{ACSPark,AMYangjl2020,CPSYJBAFM}. Based on the spin-flop transition field and the saturation field in two different directions, we estimate the interlayer coupling to be 0.16 meV and the on-site anisotropy energy to be 0.0045 meV, which are found to be very close to the neutron scattering measurement results~\cite{CPSPRB2020}. This weak anisotropy energy results in similar spin-canted states for magnetic fields applied along both directions, as evidenced by nearly identical magnetization in both cases.

\begin{figure*}
\includegraphics[width =1\linewidth]{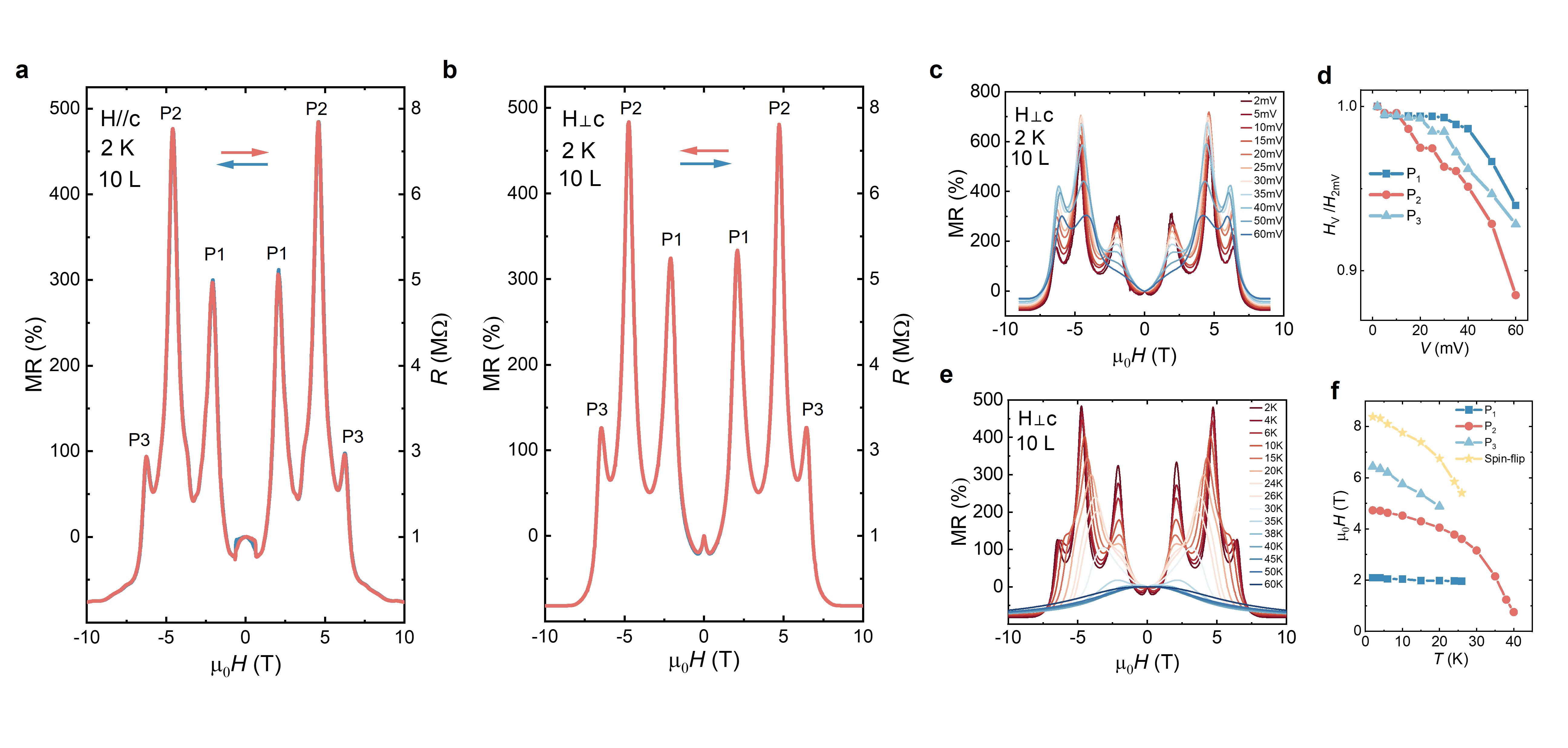}
\caption{\textbf{MR oscillations in the vertical junction of 10-layer CrPS$_4$.} \textbf{a}. Magnetic field dependence of resistance measured with the field parallel to the $c$-axis, obtained with a constant AC bias voltage of 2 mV at 2 K.  Clear MR peaks are observed, along with the spin-flop transition at approximately $\pm$ 0.6 T. \textbf{b}. MR measured with the magnetic field perpendicular to the $c$-axis. The persistence of MR oscillations indicates their origin from CrPS$_4$ rather than the graphene electrodes. \textbf{c}. MR measured with different DC bias voltages. The MR peaks gradually shift to lower fields with increasing voltage, as summarized in \textbf{d}. \textbf{e}. MR at different temperatures measured with a AC bias voltage of 2 mV. The peak positions shift to lower fields as the temperature increases, as summarized in \textbf{f}.} 
\end{figure*}

To investigate the electronic transport properties of few-layer CrPS$_4$, we fabricated vertical junctions of graphene/CrPS$_4$/graphene as illustrated in Fig. 1d (see Method for details of fabrications). Fig. 1e presents the $I-V$ curves of vertical junctions comprising multi-layer CrPS$_4$ at 2 K. Some thicker devices exhibit clearly non-linear $IV$ behavior with ultra-high resistivity at zero voltage bias. Devices exhibiting relatively linear $IV$ behavior also demonstrate high resistivity, typically over 10$^5$ $\Omega$ $\cdot$ cm. These findings indicate the insulating behavior of our devices at low temperatures, consistent with previous reports on both bulk and thin flakes of CrPS$_4$~\cite{CPSJAP, CPSNPJ, CPSAlberto,CPSAlberto2023,CPSCJH,CPSvanWees, CPS2024}. We also fabricated field effect transistor to measure the longitudinal transport (see supplementary information note 1), the resistance at zero gate voltage is very large and beyond our measurement capability. The insulating state is further corroborated by temperature-dependent resistance measurements, as depicted in Fig. 1f. The resistance increases by approximately two orders of magnitude as the device is cooled from room temperature to 50 K. Analysis of the Arrhenius plot (insert in Fig. 1f) reveals a thermal activation energy of 17.8 meV. At lower temperatures, the deviation from thermal activation transport behavior suggests electron transport occurs via tunneling or hopping processes, which will be discussed in detail later. Notably, a distinct feature is observed at the Néel temperature in the $R$ vs $T$ curve of all devices, confirming the consistent quality of our atomically thin devices and bulk samples.

Having established that our devices exhibit insulating behavior at low temperatures, we proceeded to investigate their electronic transport behavior under the influence of a magnetic field. Fig. 2a illustrates the MR of a 10-layer device, defined as $MR$ = ($R$$_H$ - $R$$_0$)/$R$$_0$, when a magnetic field is applied parallel to the $c$-axis at 2 K. In the low magnetic field region, resistance undergoes a sudden change at approximately $\pm$ 0.6 Tesla, closely matching the spin-flop transition field of bulk CrPS$_4$. In the high magnetic field region, resistance saturates at around 8 Tesla, aligning with the magnetic field at which the magnetization saturates, known as the spin-flip transition field. These observations underscore the correlation between the electronic transport behavior of few-layer vertical junctions and the magnetic states of CrPS$_4$.

\begin{figure*}
\includegraphics[width =0.9\linewidth]{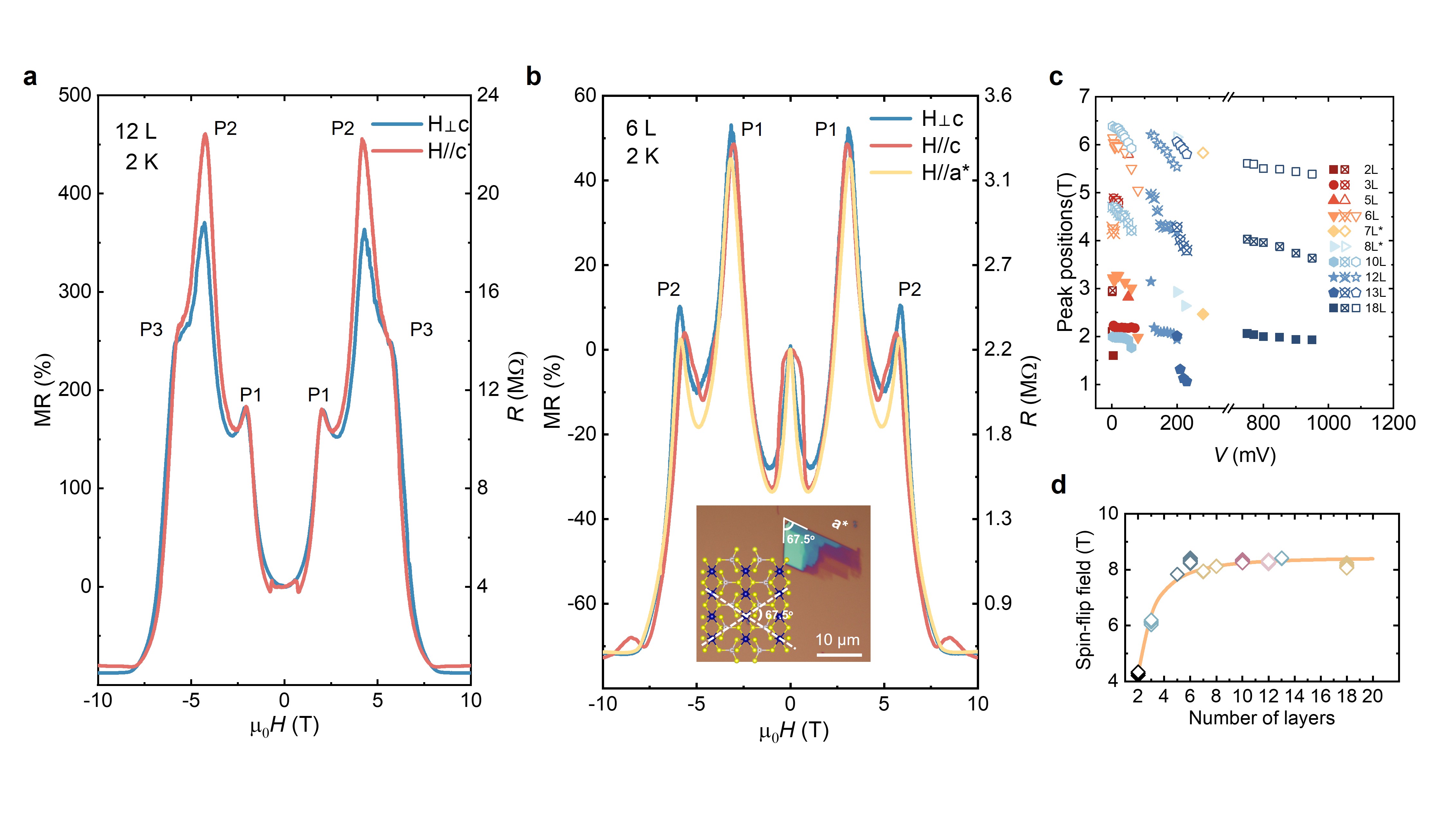}
\caption{\textbf{Statistics of MR oscillations in few-layer CrPS$_4$ devices.} \textbf{a}. MR of a 12-layer device measured with a DC bias voltage of 0.15 V. Despite being more resistive than the 10-layer device, MR peaks persist in both magnetic field orientations. \textbf{b}. MR of a 6-layer device measured with a AC bias voltage of 10 mV. The inset shows the optical image of the flake, with the vertical junction located at the black point. The flake exhibits two sharp edges intersecting at an angle of 67.5$^o$. Magnetic fields are applied parallel to the $c$-axis, $ab$-plane, and $a^*$-axis. \textbf{c}. Summary of MR peak positions for all few-layer devices. Here the 7 and 8 layer devices use NbSe$_2$ as electrodes. \textbf{d}. Spin-flip transition fields determined from in-plane MR measurements at 2 K, with the solid line representing calculated results based on parameters derived from bulk magnetization data. The layer number is determined using a combination of atomic force microscopy and optical contrast, with approximate $\pm1$ layer error for thick CrPS$_4$ flakes.} 
\end{figure*}

MR oscillations are prominently observed in the intermediate magnetic field region between the spin-flop transition and the spin-flip transition. The MR exhibits three distinct peaks at around 2 T, 4.7 T, and 6.4 T, labeled as P1, P2, and P3, respectively. These peaks are evident in both magnetic field sweep directions and exhibit symmetry with respect to zero magnetic field. It is noteworthy that the magnetization of bulk CrPS$_4$ monotonically changes with the magnetic field, as previously demonstrated in Fig. 1c. Similarly, the magnetization of few-layer flakes would also exhibit monotonic changes when considering the total magnetic energy of the system (see supplementary information Note 4 and supplementary Fig. 15b), a behavior experimentally verified in various few layer A-type antiferromagnets~\cite{CrCl3NN, PRXyeyu}. In this context, the observation of MR oscillations in CrPS$_4$ vertical junctions is unexpected and stimulates further investigation.

One plausible explanation for these MR oscillations is a change in the density of states of the graphene electrodes induced by the perpendicular magnetic field. To explore this possibility, we conducted MR measurements with a magnetic field applied perpendicular to the $c$-axis, and the corresponding data is presented in Fig. 2b. In the low magnetic field region, the MR exhibited a smooth variation, as no spin-flop transition was anticipated when the magnetic field was perpendicular to the easy axis. In the high magnetic field region, the MR saturated at around 8.2 Tesla, close to the bulk spin-flip transition field. Interestingly, in the intermediate magnetic field region, the MR oscillations remained largely unchanged in terms of both peak amplitudes and positions. 

An alternative explanation is that the MR oscillations might originate from new states in the graphene electrodes induced by the proximity effect with magnet CrPS$_4$, persisting even when the magnetic field is perpendicular to $c$-axis. To test this hypothesis, We conducted two control experiments. First, we fabricated a device with the structure Graphene/hBN/CrPS$_4$/hBN/Graphene. The inclusion of thin hBN layers between the graphene electrodes and CrPS$_4$ is intended to largely reducing the wavefunction overlap between them, thus eliminate any potential new states in graphene induced by proximity effect. Second, we constructed another type of device with the structure NbSe$_2$/CrPS$_4$/NbSe$_2$, where 2D metal NbSe$_2$ served as the electrodes, eliminating graphene entirely. As detailed in the appendix, MR oscillations were clearly observed in both types of control devices.

These experiments provide robust evidence that the MR oscillations originated from CrPS$_4$ itself, rather than from the graphene electrodes. In the following we focus on the devices of Graphene/CrPS$_4$/Graphene. Fig. 2c illustrates the MR measurements of 10 layer device conducted under different voltage biases at 2 K. The MR oscillations persisted across all measured biases, albeit with decreasing amplitudes as the bias increased. At a bias of 60 mV, the identification of the first peaks became less straightforward. Interestingly, the positions of the MR peaks shifted to lower magnetic fields with increasing bias, as summarized in Fig. 2d.

We further investigated the temperature dependence of the MR behavior. As shown in Fig. 2e, the amplitudes of the MR oscillations decreased with increasing temperature, ultimately vanishing at around 40 K, close to the Néel temperature of CrPS$_4$. In addition to affecting the amplitude of the MR oscillations, increasing temperature also shifted all three peaks to lower magnetic field, as summarized in Fig. 2f. This temperature dependence contrasts with that of typical quantum MR oscillations observed in conducting systems, such as Shubnikov–de Haas oscillations and the Aharonov-Bohm effect, where MR peak positions remain independent of temperature. The similar trend in temperature dependence between the MR oscillations and the spin-flip transition suggests a connection to the canted magnetic states of CrPS$_4$.

Before delving deeper into the origin of MR oscillations, we investigated this behavior in devices with varying thicknesses. Fig. 3a presents the MR of 12-layer device, characterized by a clearly non-linear $IV$ curve, indicative of even more insulating compared to 10-layer device. Three MR peaks are distinctly observed in both perpendicular and parallel magnetic field configurations, with the middle peak (P2) dominating in comparison to 10-layer device. Turning to the results of the 6-layer device, as shown in Fig. 3b, only two MR peaks are observed in most measurements, with an additional small MR peak emerging when the voltage is less than 5 mV (see Supplementary Fig. 9). This specific flake has clear crystal crossed with angle of $67.5^o$, meaning the crystal edge is a crystal axis~\cite{CPSACS2017, ACSPark} as indicated in the inset  of Fig. 3b. We applied the magnetic field along this axis and the measured MR is almost the same as the result of randomly applied in-plane field.  

We have measured total 11 multilayer devices and Figs. 3c/3d summarize the featured magnetic fields of all devices (refer supplementary information for additional electronic transport data of devices). Fig. 3c illustrates MR peak positions measured under various voltages at 2 K, where each color represents data from one device and each symbol denoting the same peak in that device. It is evident that all MR peaks shift to lower magnetic fields with increasing voltage, and they can be roughly divided into four groups. Fig. 3d depicts the thickness dependence of the spin-flip field. The dots represent data obtained from the saturation of magnetoconductance in the parallel field configuration, particularly the bi-layer and three-layer device reveals a clear decrease. This reduction aligns with expectations, as the effective interlayer coupling energy per layer decreases. This phenomenon can be fully elucidated using the antiferromagnetic linear chain model, as discussed in the supplementary materials. 

Expanding our investigation to the atomically thin limit, we studied mono-layer CrPS$_4$, which has been identified as a ferromagnet\cite{ACSPark}. Its zero-field temperature dependence of resistance monotonic increase deviated at around 38 K (the arrow in Fig. 4a), which is around the same as the Néel temperature of bulk CrPS$_4$ ~\cite{CPSACS2017,CPS2024,ACSPark}. Fig. 4b presents measurement results with a magnetic field perpendicular to the $c$-axis, where no MR oscillations are observed in this monolayer device. The MR is around zero at 2 K and becomes negative as temperature rises. The MR amplitude increases as temperature rises and reaches maximum at around Curie temperature, exhibiting a clear triangle shape when plotted as function of temperature in a 2D format (see supplementary materials). This behavior closely resembles that observed in vertical junctions of other 2D ferromagnetic semiconductors, such as CrBr$_3$ ~\cite{CrBr3NC} and mono-layer CrSBr ~\cite{CrSBrAM}. In contrast to antiferromagnets, no spin-canted states are expected in ferromagnets. The absence of MR oscillations in the mono-layer device thus further confirms that the oscillatory behavior is related to spin-canted states.


\begin{figure}
\includegraphics[width =1\linewidth]{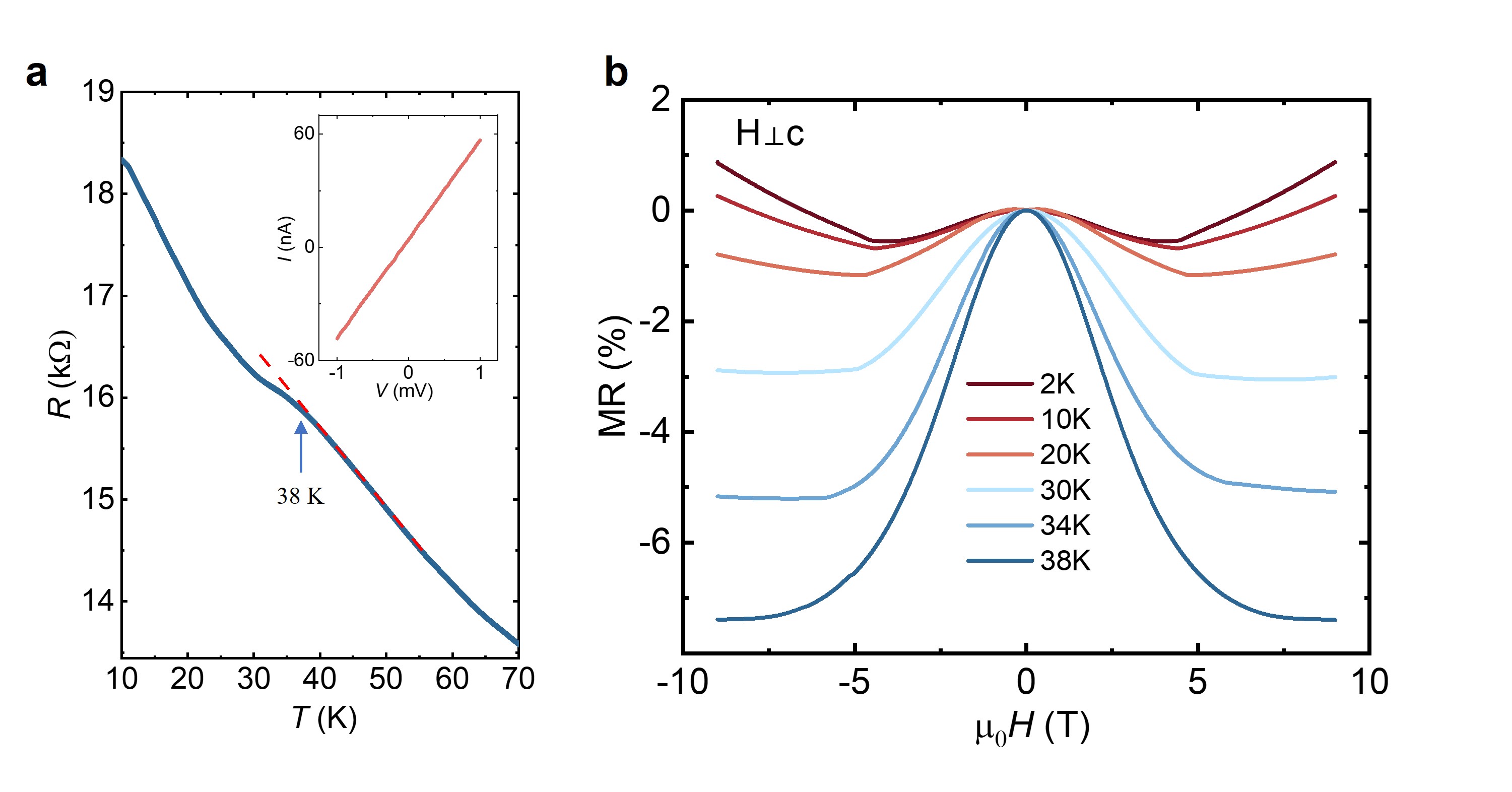}
\caption{\textbf{Absence of MR oscillations in ferromagnetic monolayer CrPS$_4$ device.} \textbf{a}. Temperature dependence of resistance at zero field. The inset shows the $IV$ curve of this vertical junction. \textbf{b}. MR at different temperatures, measured with a magnetic field perpendicular to the $c$-axis, no MR oscillation is observed.} 
\end{figure}

\section*{III.	Discussions}
In summary of the aforementioned experimental investigations across various devices, no MR oscillations are observed in monolayer device due to its ferromagnetic nature. Conversely, for antiferromagnetic few-layer devices, MR oscillations occur within the region between spin-flop and spin-flip transitions, with peak positions shifting in tandem with the spin-flip transition. These findings strongly suggest that MR oscillations are intricately linked to the magnetic canted states induced by the applied magnetic field. 

To gain deeper insights into the specifics of these canted states in multi-layer CrPS$_4$, we employ the antiferromagnetic linear-chain model~\cite{CrCl3NN,PRXyeyu}. Within this framework, the magnetization of each CrPS$_4$ layer is depicted as a macro-spin with an on-site anisotropy energy $K$, coupled to its nearest neighbors through the interlayer coupling $J$. The magnetic state under a certain magnetic field can be elucidated by minimizing the total magnetic energy of the system (detailed in the supplementary note 3 and 4). This model can nicely describe the layer number dependence of spin-flip field shown in Fig. 3d, where the solid orange line represents the calculated results with parameters derived from bulk magnetization measurements. Our analysis reveals that while the total magnetization monotonically increases (Supplementary Fig. 15b), the canting angle undergoes non-monotonic changes as a function of the magnetic field, for example, the canting angle of layer 2 and 5 in the 6-layer CrPS$_4$ (supplementary Fig. 15a).

Despite our insights into the magnetic states through the antiferromagnetic linear-chain model, establishing the link between the magnetic states and electronic transport proves elusive. Our calculations of  MR based on the tunneling model used in CrCl$_3$ (see details in the supplementary note 4) reveal a monotonic behavior, with no oscillations observed. This points out the importance of evaluating the predominant transport mechanism in vertical junctions of CrPS$_4$. Our analysis of the saturation of low-temperature resistance in the Arrhenius plot (see the inset of Fig. 1f) suggests that thermally activated electrons are not the predominant carriers within the examined temperature range. These imply the presence of states inside the gap of CrPS$_4$. This conclusion is also consistent with previous report on field-effect transistors of CrPS$_4$\cite{CPSAlberto}, where only monotonic MR is observed when the Fermi level is tuned into the bands of CrPS$_4$, as the transport is not mediated by the in-gap states.

The nature of these in-gap states remains elusive. Previous observations of quantum MR oscillations in Kondo insulator YbB$_1$$_2$ and topological insulator WTe$_2$ suggest an unconventional Fermi surface inside the insulating gap~\cite{YbB12,WTe2}, yet their origins remain inconclusive. These oscillations, resembling Shubnikov-de Haas oscillations in metals, differ from those in our device, and the band gap opening mechanism in these systems also differs from CrPS$_4$. Besides the possibility of these exotic states, defects offer a trivial explanation for in-gap states in our devices. At high temperatures, hopping between defect states leads to thermally active behavior, with the activation barrier determined by the energy spacing between these states~\cite{MottBook,HoppingNC}. This scenario elucidates the ultra-small thermal activation energy (17.8 meV) obtained from the Arrhenius plot, notably lower than the band gap (1.3 eV) of CrPS$_4$. With these defect states, variable-range hopping typically dominates at low temperatures in standard longitudinal devices~\cite{MottBook, Shklovskii}. However, given the small distance between the two graphene electrodes of our vertical junctions (few nanometers), direct application of classic variable-range hopping theory is untenable. Nevertheless, we can conceptualize the transport process in a similar manner, where electrons tunnel from one defect state to another suitable defect state, either within the same layer or in neighboring layers, until they reach the graphene electrode on the opposite side. The average hopping length is determined not only by the sample thickness but also by the defect density and the energy difference between these defects, potentially explaining why some thicker devices exhibit smaller resistance than thinner one. 

To further investigate the correlation between the electronic transport properties and the magnetic states in CrPS$_4$, we calculated the band structure of CrPS$_4$ containing sulfur (S) vacancies, a common type of defect in 2D transition metal chalcogenides. As detailed in the supplementary materials, our calculations reveal localized states within the band gap of CrPS$_4$. Notably, the S vacancy is spin-polarized, aligning with the spin direction of the Cr atoms in the layer. In the antiferromagnetic ground state, the substantial energy difference between spin-polarized defect states in adjacent layers (when considering spin conserved hopping) necessitates electron hopping to more distant layers, which increases the hopping length and, consequently, the resistance. When an external magnetic field is applied, spin canting reduces the energy difference between neighboring layers, facilitating hopping and thereby reducing the hopping length and the resistance. Concurrently, within conventional magneto-hopping-conduction theory~\cite{Shklovskii}, the magnetic field compresses the wavefunction of the localized states, decreasing the probability of hopping and thus increasing the hopping length and resistance~\cite{VRHBfield}. The interplay between these competing effects—spin canting and wavefunction contraction—accounts for the MR peaks observed in our measurements. Additionally, the application of an electric field can effectively reduce the energy difference between defect states, leading to a decrease in the hopping length~\cite{VRHEfield0,VRHEfield2,VRHEfield3}. This dependence on electric field strength modulates the competition between the spin-canting effect and the magnetic field effect, resulting in shifts in the positions of the MR peaks.

The interlayer hopping of spin-polarized defect states in CrPS$_4$ offers a plausible framework to quantitatively explain our key findings. We employed principles of conventional variable range hopping theory to develop a toy model, the spin selected interlayer hopping model as detailed in the supplementary materials, to calculate the MR. This model successfully captures the MR peaks and their shifts under an electric field. \comment{However, it does not accurately predict the peak positions observed in the experimental results.} The exact MR oscillations behavior, such as the MR magnitudes and peak positions, are influenced by the intricate behavior of the localized wavefunction in response to the magnetic field, as well as detailed calculations of hopping resistance. It is important to note that a comprehensive theory of interlayer hopping involving spin-polarized states in 2+1 dimensions—considering the strong anisotropy along the $c$-axis in our system—has yet to be developed. Further effort are needed to clarify the exact nature of the in-gap states and to establish a complete hopping conduction framework for such systems if spin-polarized defect states are indeed found to be responsible for the observed MR oscillations. \comment{At the same time, we notice that the position of MR peaks can be more precisely described when we consider the interference between spin Berry phases. In the simplified scenario,} the spin can follow two distinct evolutionary paths in a single hopping process due to the coexistence of spin-up and spin-down channels. Different spin Berry phases emerge for each of these cyclic quantum evolution paths \cite{spinphase, ZYNP}. The interference between these two paths, which is strongly dependent on the spin-canting configuration, further affects the overall hopping probability \comment{as detailed in the supplementary materials. However, more efforts are needed to further support the existence of spin Berry phase coherence and investigate its precise origin.}


\section*{IV.	Summary}
In conclusion, our study reveals the presence of robust MR oscillations in vertical junctions of few-layer CrPS$_4$. These oscillations persist regardless of the direction of the applied magnetic field, whether parallel or perpendicular to the c-axis. They are observed across a spectrum of samples, ranging from highly insulating (exhibiting clearly non-linear $I-V$ curves with unmeasurable resistance at zero voltage bias) to those with relatively lower resistance (still in the order of Megaohm), as long as the device comprises multilayer CrPS$_4$. Moreover, the MR peaks exhibit a gradual shift to lower magnetic fields as the temperature increases, ultimately disappearing above the Néel temperature, mirroring the behavior of the spin-flip transition. This correlation underscores the connection between these MR oscillations and the canted magnetic state in CrPS$_4$. While further investigations are necessary to precisely determine \comment{the proposed spin-polarized defect states and possible coherence between spin Berry phase}, our findings highlight CrPS$_4$ as a unique example of MR oscillations, a phenomenon rarely observed in insulating systems. This work underscores the potential of 2D van der Waals magnets not only for spintronics applications but also as intriguing platforms for exploring novel physics. 

We notice similar works \cite{ZYNP, AlbertoarXiv} after submission of this manuscript. 

\section*{Appendix}
\textbf{Bulk crystal growth and characterizations}: CrPS$_4$ crystals were grown using the Chemical Vapor Transport method, following established protocols\cite{CPSGrowth}. Chromium, red phosphorus, and sulfur powders were measured stoichiometrically (Cr: P: S = 1:1:4) with 5 percentage additional sulfur added as transport agent. The precursors were mixed and sealed into quartz ampoules under vacuum (10$^{-2}$ Pa), followed by loading into a two-zone furnace. The hot and cold ends of the ampoules were kept at 680 °C and 600 °C, respectively for 8 days. At the end of the growth process, the furnace was turned off for natural cooling and the CrPS$_4$ crystals could be collected from the cold zone of the ampoule. X-ray diffraction were performed to determine the structure of the grown crystals. One crystal weighing 1.13 mg was utilized for magnetization measurements performed in a PPMS VSM magnetometer (Quantum Design), with the magnetic field oriented parallel or perpendicular to the crystallographic $c$-axis. 

\textbf{Vertical junction fabrication and transport measurements}: Atomically thin CrPS$_4$ flakes were obtained through mechanical exfoliation from bulk crystals. The vertical junctions of graphene/CrPS$_4$/graphene were assembled using a standard pick-up technique with stamps of PDMS/PC. To ensure the high quality of the vertical junctions, exfoliation and assembly were conducted within a glove box filled with nitrogen gas, and the junctions were encapsulated with hBN. Standard nanofabrication processes, including electron beam lithography, reactive ion etching, and electron beam evaporation (10 nm/50 nm Cr/Au), were employed to make contacts to the graphene electrodes. Electronic transport measurements were conducted in a cryostat from Cryogenic or Physical Property Measurement System (PPMS) from Quantum Design, using a Keithley 2400 and SR830 lock-in amplifier. MR of multi-layer devices were measured with constant bias voltage, and the mono-layer device was measured with constant current of 40 nA. Frequency of 17.773 Hz was used in AC measurements.

\textbf{Electronic transport data of vertical junctions with different electrodes:}
As disscused in the main text, one possible explanation of MR oscillations is that they might originate from new states in the graphene electrodes induced by the proximity effect with magnet CrPS$_4$. To address this, We conducted two control experiments. First, we fabricated a device with the structure Graphene/hBN/CrPS$_4$/hBN/Graphene. The proximity effect typically involves the expansion of the wavefunction from one material into another, resulting in new properties in the latter. This wavefunction expansion decreases exponentially with increasing distance between the materials. For instance, ferromagnetism in graphene induced by proximity to a yttrium iron garnet (YIG) substrate disappears when a very thin Al$_2$O$_3$ layer is inserted between them\cite{proximity}. In our device, the thin hBN layer inserted between CrPS$_4$ and graphene would eliminate any potential new states in graphene induced by proximity to CrPS$_4$. As shown in Fig. 5 , MR oscillations were still clearly observed when magnetic field is applied parallel or perpendicular to the $c$-axis of the device, providing evidence that these oscillations originate from CrPS$_4$ itself.

\begin{figure*}
\centering
\includegraphics[width =0.9\linewidth]{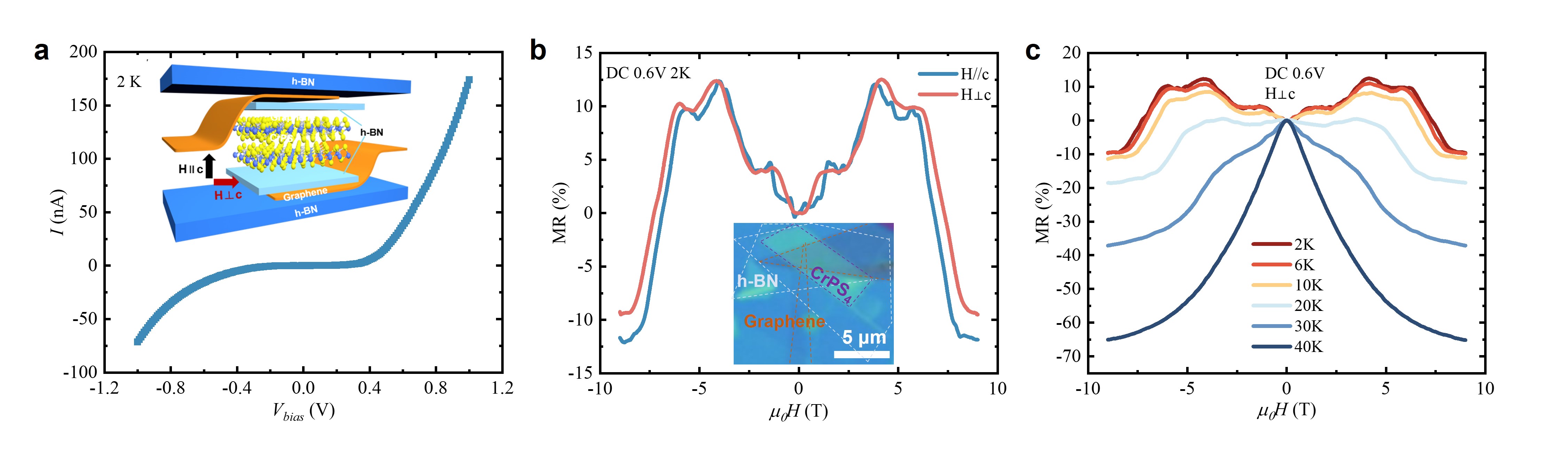}
\caption{\textbf{MR oscillations in vertical junction of Graphene/hBN/CrPS$_4$/hBN/Graphene}. a. IV curve of the device at 2 K and zero field, the difference thickness of inserted hBN results in the asymmetry of the device. The inset shows schematic of vertical junction Graphene/hBN/CrPS4/hBN/Graphene, where thin hBN was inserted between graphene electrodes and CrPS$_4$. b. MR oscillations at 2 K with magnetic field applied parallel and perpendicular to the $c$-axis. The measurement is done with constant DC voltage of 0.6 V. The inset show the optical image of the device. c. MR at different temperatures, the magnetic field is applied perpendicular to $c$-axis.}
\end{figure*}

\begin{figure*}
\centering
\includegraphics[width =0.9\linewidth]{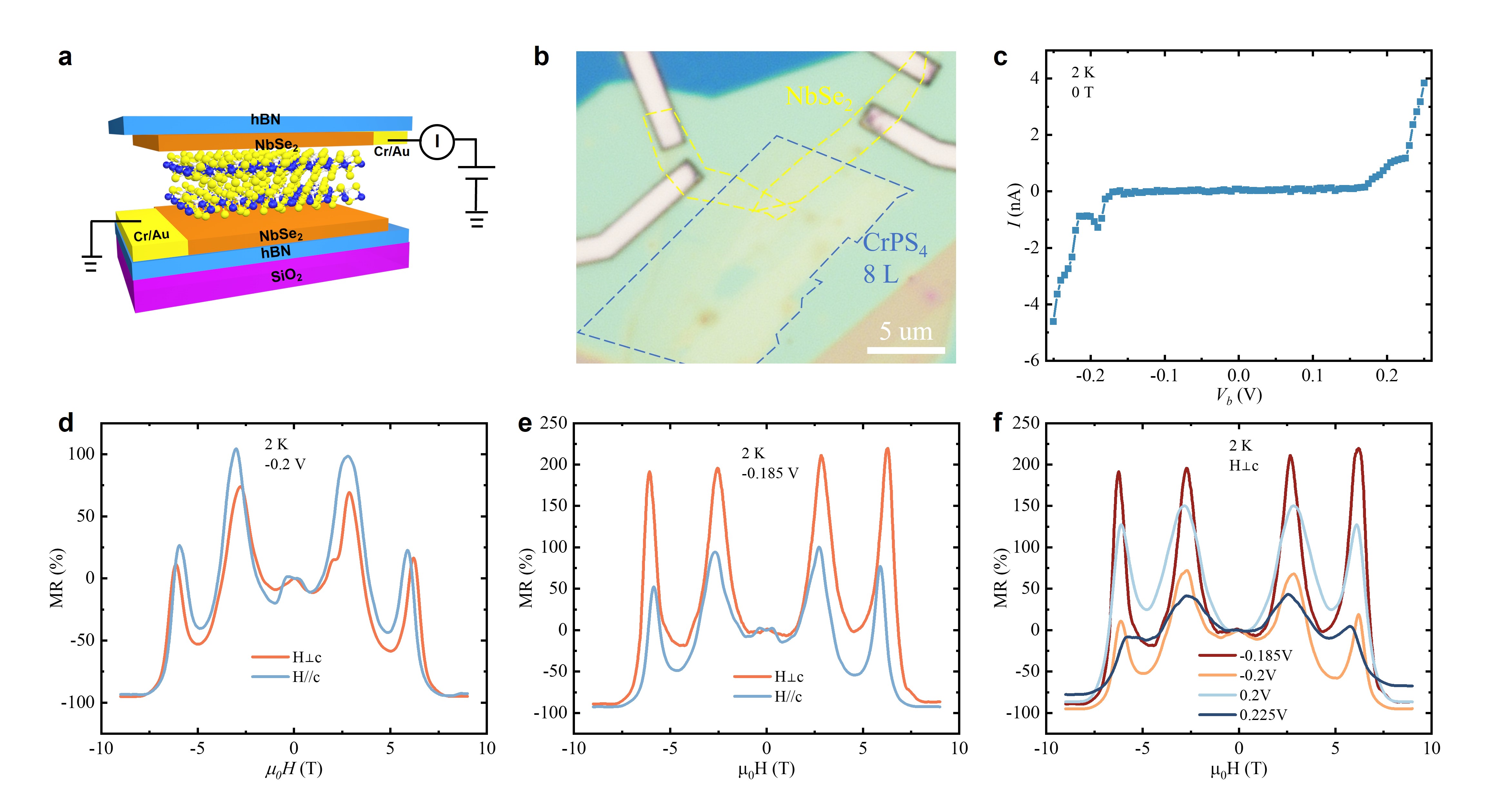}
\caption{\textbf{MR oscillations in vertical junction of NbSe$_2$/CrPS$_4$/NbSe$_2$}. a. Schematic of vertical junction NbSe$_2$/CrPS$_4$/NbSe$_2$. b. Optical image of device NS1. c. IV curve measure at 2 K at zero magnetic field. d. MR oscillation at 2 K with magnetic field parallel and perpendicular to the c-axis. Applied DC bias voltage is -0.2 V. e. MR oscillation at 2 K with DC bias voltage is -0.185 V. f. MR oscillation at 2 K with different DC bias voltage, magnetic field is perpendicular to c-axis.}
\end{figure*}

\begin{figure*}
\centering
\includegraphics[width = 0.9\linewidth]{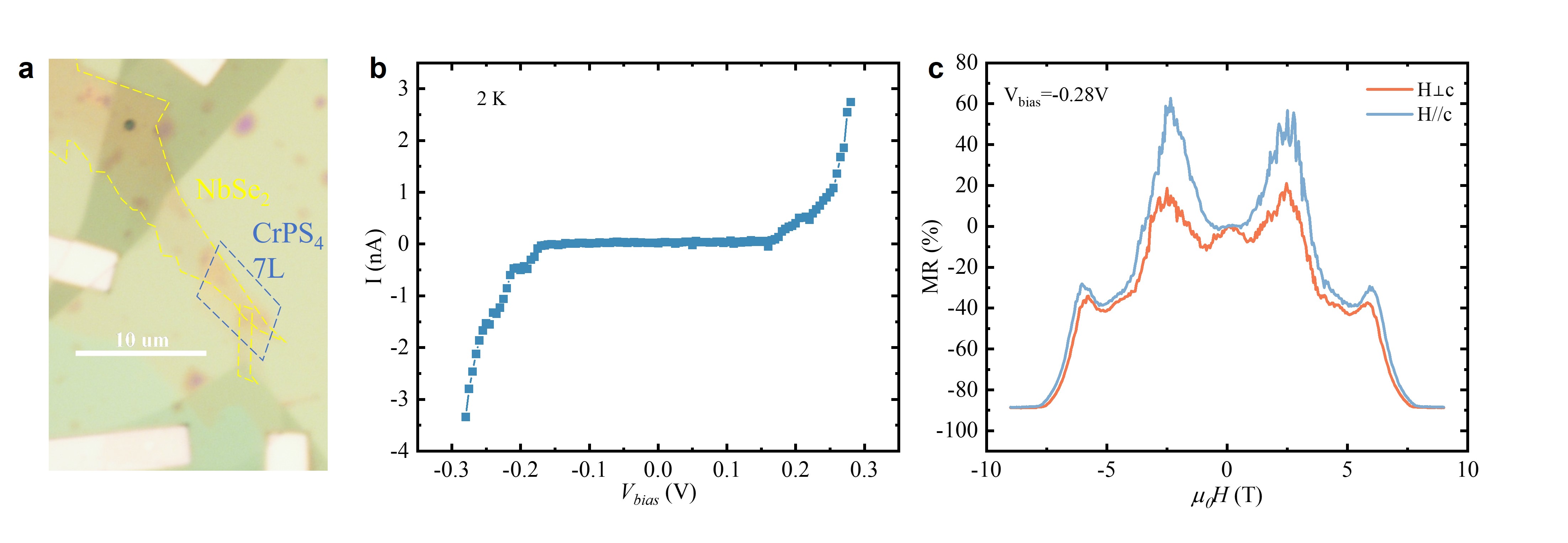}
\caption{\textbf{MR oscillations in another vertical junction of NbSe$_2$/CrPS$_4$/NbSe$_2$}. a. Optical image of device NS2. b. IV curve measure at 2 K at zero magnetic field. c. MR oscillation at 2 K with magnetic field parallel and perpendicular to the $c$-axis. Applied DC bias voltage is -0.28 V.} 
\end{figure*}

Second, we constructed another type of device with the structure NbSe$_2$/CrPS$_4$/NbSe$_2$ as illustrated in Fig. 6a. We use 2D metal NbSe$_2$ as the electrodes and omit graphene in this type of device. The fabrication process for this device was similar to that described in the main text for the graphene/CrPS$_4$/graphene device. To avoid the oxidation of NbSe$_2$, the device was immediately loaded into the e-beam evaporator’s vacuum chamber after etching the hBN, with NbSe$_2$ exposure to air minimized to less than one minute. Fig. 6 shows the results for one such device, NS1. The non-linear IV curve at 2 K and 0 T is presented in Fig. 6c. We observed features around ±0.2 V in the IV curve, which were also seen in another NbSe$_2$/CrPS$_4$/NbSe$_2$ device NS2, though their origin is unclear. Figs. 6d and e show the MR at 2 K measured with biases of -0.2 V and -0.185 V, respectively, where MR oscillations are evident. Fig. 6f shows the MR at 2 K measured with both positive and negative biases, with the magnetic field applied perpendicular to the $c$-axis, and MR oscillations are observed in all measurements.

For another NbSe$_2$/CrPS$_4$/NbSe$_2$ device, NS2, we used graphene flakes to make contacts to NbSe$_2$, ensuring NbSe$_2$ did not need to be exposed to air during fabrication (note that graphene did not contact CrPS$_4$). The results were very similar to those for device NS1, as shown in Fig. 7. The consistent data from one Graphene/hBN/CrPS4/hBN/Graphene device and two NbSe$_2$/CrPS$_4$/NbSe$_2$ devices provide robust evidence that the MR oscillations originate from the magnetic semiconductor CrPS$_4$ itself.

\section*{Acknowledgement}
This work is financially supported by National Natural Science Foundation of China (Grants no. 12374121 , 12304232 and 12274090), Shaanxi Fundamental Science Research Project for Mathematics and Physics (22JSY026), the Fundamental Research Funds for the Central Universities and the Natural Science Foundation of Shanghai (Grant no.22ZR1406300). K.W. and T.T. acknowledge support from the JSPS KAKENHI (Grant Numbers 20H00354 and 23H02052) and World Premier International Research Center Initiative (WPI), MEXT, Japan.

\bibliographystyle{mynaturemag}
\bibliography{biblio.bib}

\end{document}


\title{Magnetoresistance oscillations in vertical junctions of 2D antiferromagnetic semiconductor CrPS$_4$}
\author{Pengyuan Shi}
\author{Xiaoyu Wang}
\author{Lihao Zhang}
\affiliation{\MOE}
\author{Wenqin Song}
\affiliation{\FD}
\author{Kunlin Yang}
\affiliation{\FD}
\author{Shuxi Wang}
\author{Ruisheng Zhang}
\affiliation{\MOE}
\author{Liangliang Zhang}
\affiliation{\ME}
\author{Takashi Taniguchi}
\affiliation{\JAPP}
\author{Kenji Watanabe}
\affiliation{\JAP}
\author{Sen Yang}
\author{Lei Zhang}
\affiliation{\MOE}
\author{Lei Wang}
\affiliation{\SE}
\author{Wu Shi}
\affiliation{\FD}
\affiliation{\FDD}
\author{Jie Pan}
\author{Zhe Wang}
\affiliation{\MOE}

\maketitle

\titleformat{\section}[block]
  {\normalfont\large\bfseries}
  {Supplementary Note \thesection:}{0.5em}{\centering}

\renewcommand\thesection{\arabic{section}}
\titleformat{\section}[block]
  {\normalfont\large\bfseries}
  {Supplementary Note \thesection:}{0.5em}{\centering}
\makeatother

\newcommand\4{$_4$}
\renewcommand\a{{\bf a}}
\renewcommand\b{{\bf b}}
\renewcommand\c{{\bf c}}
\renewcommand\d{{\bf d}}

\section{Basic characterizations of bulk crystals and thin flakes}

Single crystals were grown using the chemical vapor transport (CVT) method, following an established protocol\cite{CPSGrowth} as described in the Appendix. Inset of supplementary Fig.1 show the optical image of single crystals. X-ray diffraction (XRD) was performed to determine the structure of the CrPS$_4$ crystals, which was measured by a Bruker D8 Advance (40 kV, 40 mA) using a Cu K$_\alpha$ ($\lambda_{\rm avg}$= 1.5418 Å). The diffraction pattern was recorded from 10° to 80° to demonstrate the consistent (00l) reflections, which shows very strong preferential orientation due to the layered structure of CrPS$_4$ crystals (supplementary Fig. 1). Energy-dispersive X-ray (EDX) spectroscopy (ZEISS GeminiSEM 500 equipped with an Oxford UltimMax100 detector) measurements were conducted on exfoliated CrPS$_4$ samples to illustrate the homogeneous atomic ratio and the uniform distributions of each element were confirmed by the EDX mapping.

\begin{figure}[H]
\centering
\includegraphics[width =0.5\linewidth]{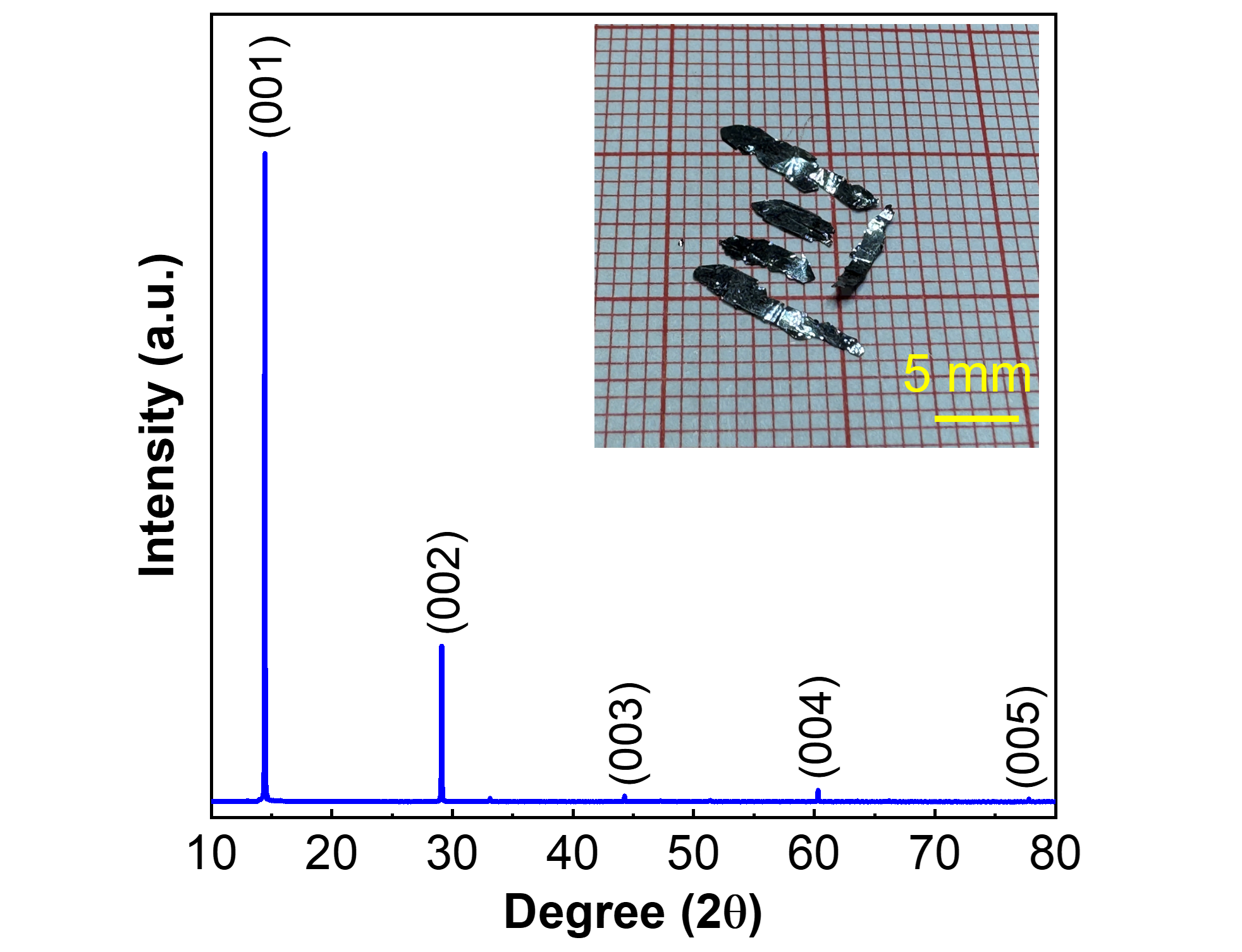}
\caption{\textbf{XRD characterization of grown single crystal. The inset shows the optical image of grown crystals.} }
\end{figure}

We conducted magnetization measurements on bulk crystal using a VSM magnetometer in a PPMS system (Quantum Design) at various temperatures. Supplementary Fig. 2 illustrates the measurement results with the magnetic field oriented parallel or perpendicular to the crystallographic $c$-axis, respectively. The spin-flop transition and spin-flip transition shift to lower field with increased temperature, as expected for antiferromagnet. The saturation magnetization at 2 K was estimated to be approximately $ \sim 3\mu_B$/Cr, consistent with previous reports.

\begin{figure}[H]
\centering
\includegraphics[width =0.9\linewidth]{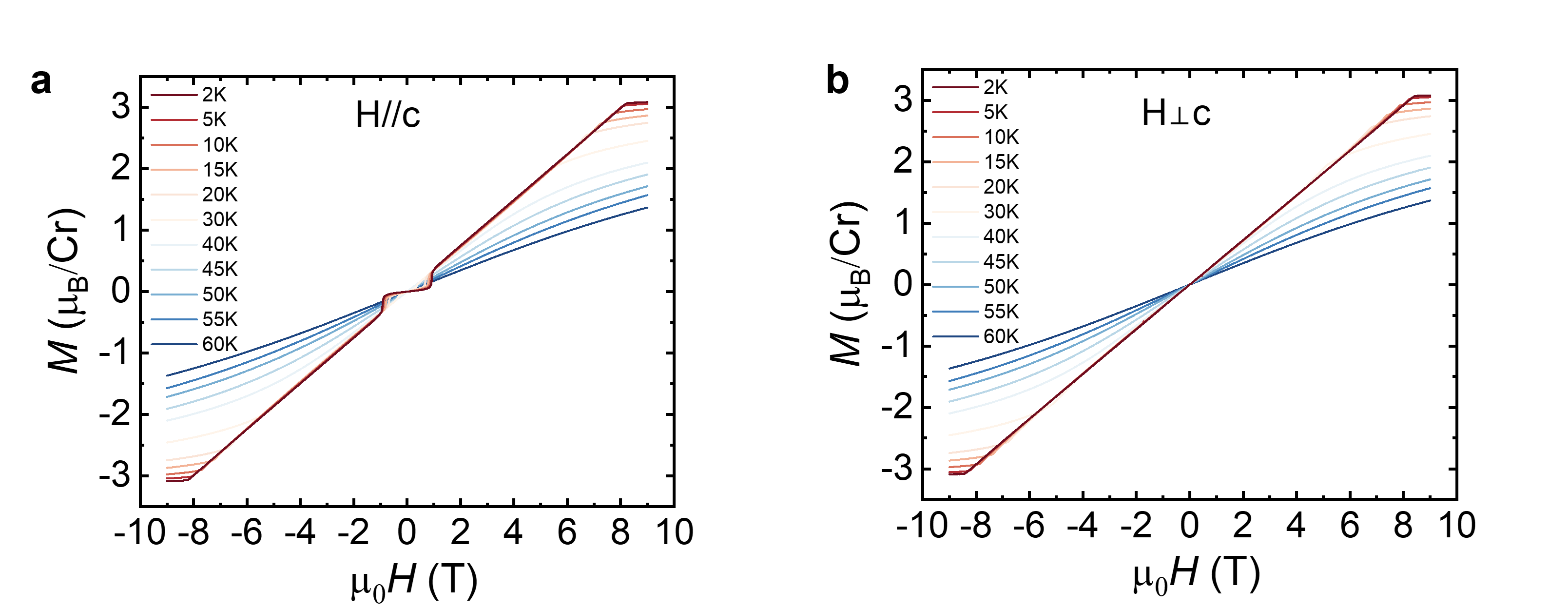}
\caption{\textbf{Magnetic field dependence of magnetization}. a. Magnetic field parallel to $c$-axis. b. Magnetic field perpendicular to $c$-axis.} 
\end{figure}

The number of layers in our CrPS$_4$ devices was initially estimated based on the optical contrast of exfoliated flakes on a 285 nm SiO$_2$/Si substrate, then was subsequently confirmed using atomic force microscopy (AFM) after assembling the heterostructure. Supplementary Fig. 3a displays the optical image of a CrPS$_4$ flake containing varying thicknesses ranging from 1 to 10 layers. Supplementary Figs. 3b and 3c present AFM images taken from different areas, further validating the layer numbers. This flake served as a standard reference for initially estimating the thickness of CrPS$_4$ prior to assembling the vertical junctions.

\begin{figure}[H]
\centering
\includegraphics[width =0.9\linewidth]{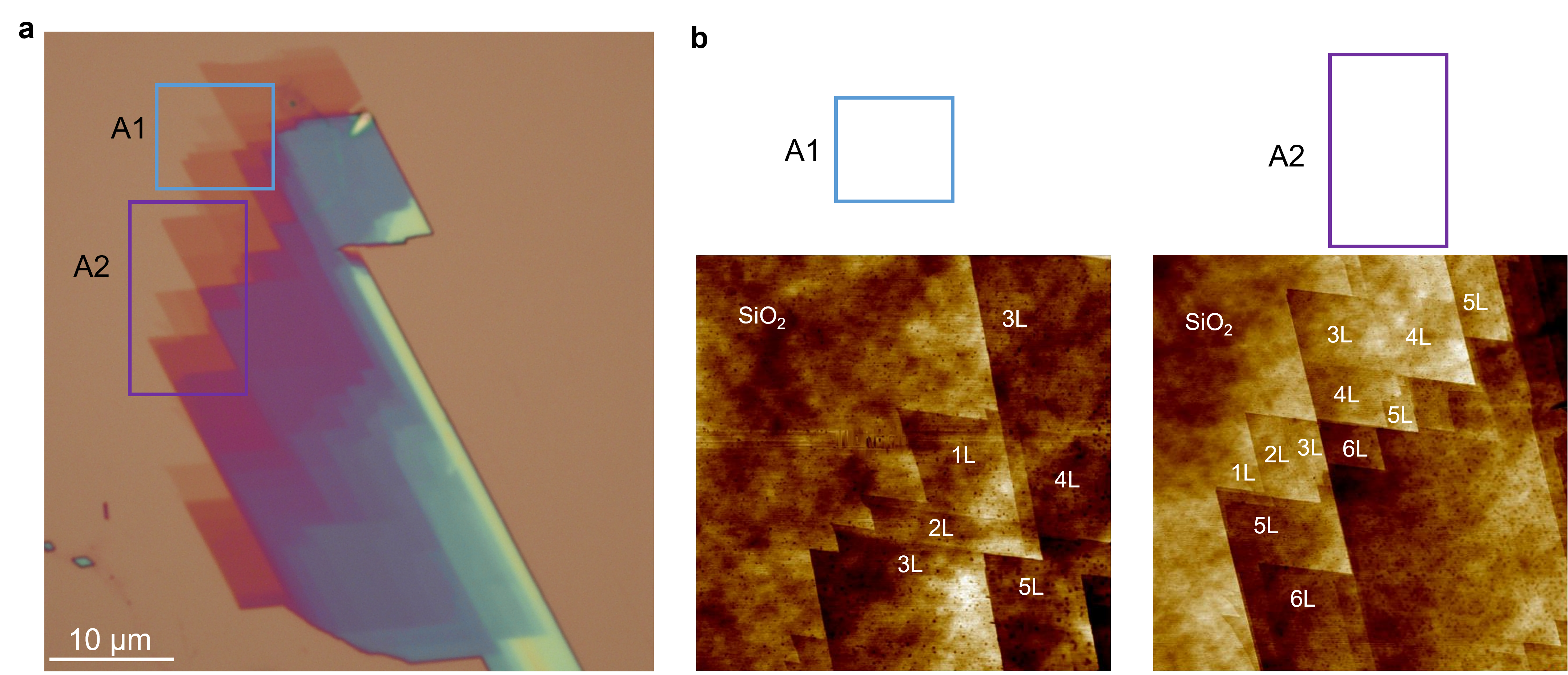}
\caption{\textbf{Images of CrPS$_4$ flakes}. a. Optical image of one flake with different number of layers. b. Atomic force microscopy image of the area indicated in a. The exact number of each part is identified by measuring its thickness} 
\end{figure}

In supplementary Table 1 and 2 we summarized basic information of all vertical junctions, including the optical images of CrPS$_4$, graphene or NbSe$_2$ electrodes, the final heterostructure and the area of vertical junctions.  
\renewcommand{\figurename}{{\bf Supplementary Table.}}

\clearpage
\newpage

\begin{figure}[H]
\addtocounter{figure}{-3}
\centering
\includegraphics[width =0.8\linewidth]{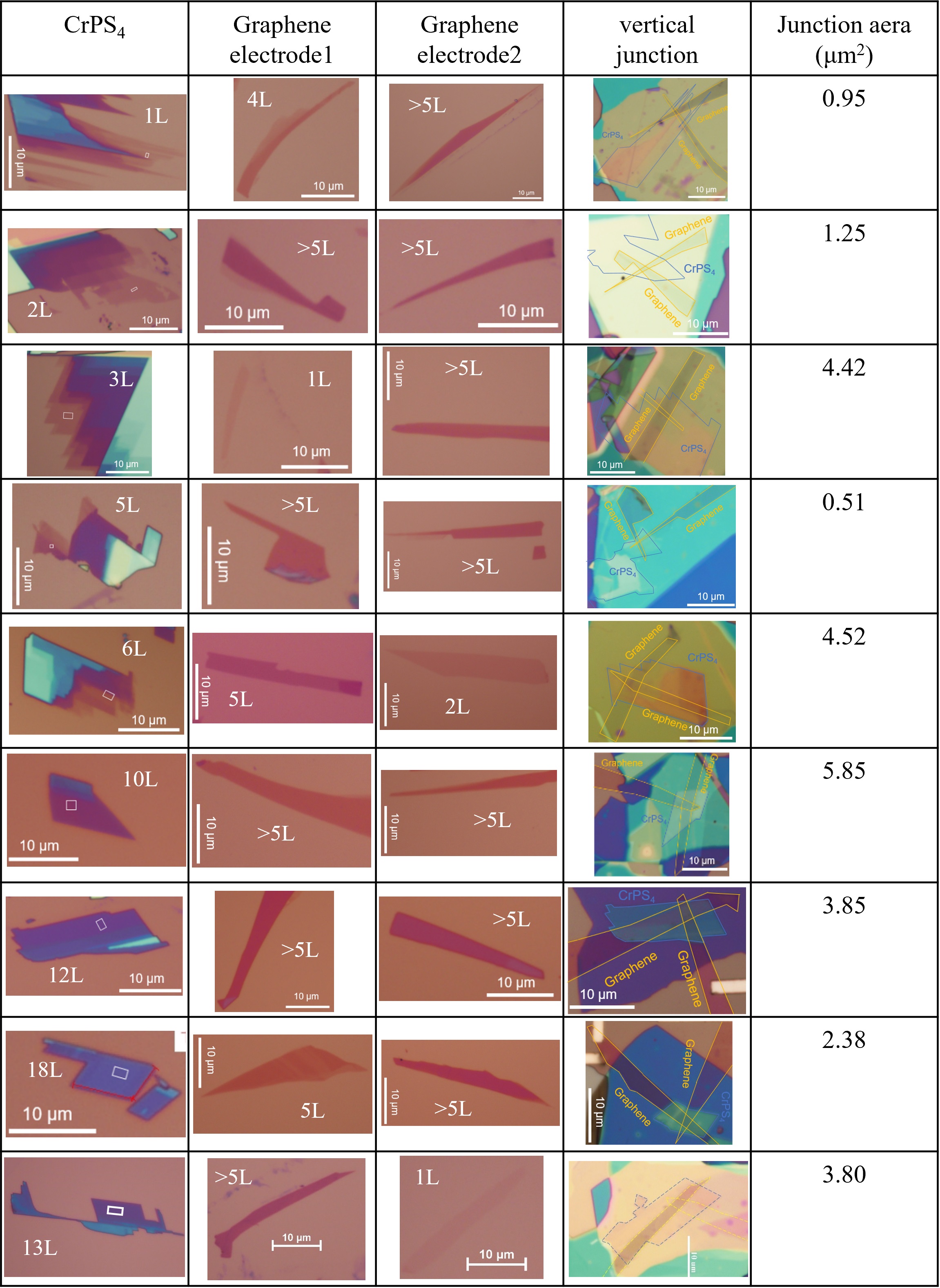}
\caption{\textbf{Images of CrPS$_4$ Devices}. Optical images of  CrPS$_4$, graphene, final heterostructures and junction area.} 
\end{figure}

\begin{figure}[H]
\centering
\includegraphics[width =0.8\linewidth]{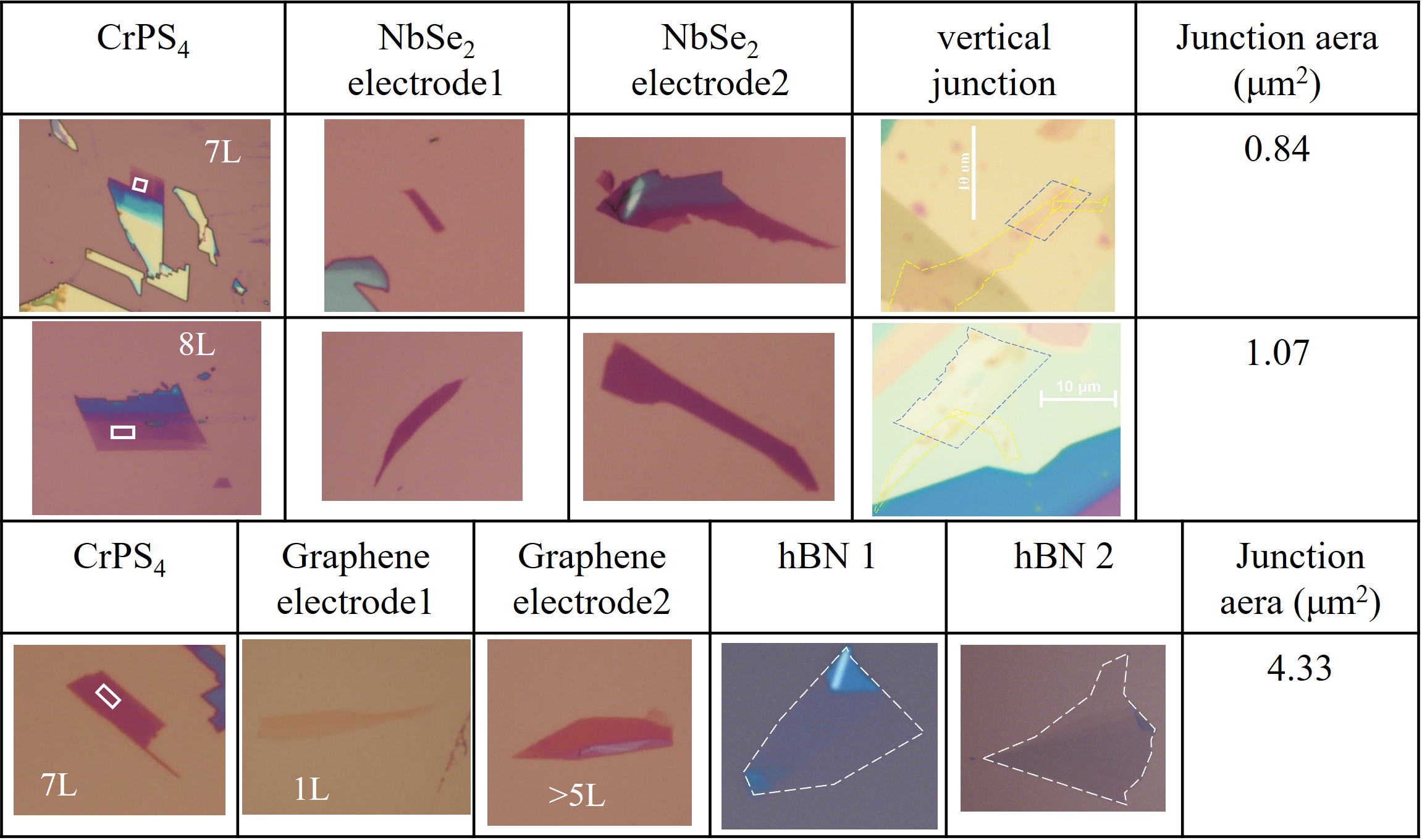}
\caption{\textbf{Images of CrPS$_4$ Devices}. Optical images of  CrPS$_4$, graphene and NbSe$_2$ electrodes, hBN, final heterostructures and junction area.} 
\end{figure}

\renewcommand{\figurename}{{\bf Supplementary Fig.}}
In addition to the vertical junctions discussed in the main text, we fabricated field-effect transistors to investigate the in-plane properties of CrPS$_4$. The fabrication process is very similar as the vertical junctions, that is, using PDMS/PC stamps to "pick-up" exfoliated hBN, CrPS$_4$ and graphene inside the glove box. Unlike the vertical junctions, the graphene strips in this device are on the same side of CrPS$_4$, close to the SiO$_2$ dielectric gate. Supplementary Fig. 4a presents the optical image of one such device of around 8 layers, where graphene strips were utilized as electrodes to make contacts with CrPS$_4$. Electronic transport measurements at 300 K are depicted in supplementary Figs. 4a and 4b. Overall, the resistance is notably high, rendering it unmeasurable when the gate voltage is below 30 V and still exceeding 1 G$\Omega$ with a gate voltage of 70 V. As temperature decreases, the resistance increases rapidly and becomes unmeasurable soon, consistent with our observations in the main text indicating that CrPS$_4$ resides in the insulating regime at low temperatures. Therefore, the in-plane transport is expected to have no significant influence on the vertical transport at low temperature.

\begin{figure}[H]
\addtocounter{figure}{1}
\centering
\includegraphics[width =0.9\linewidth]{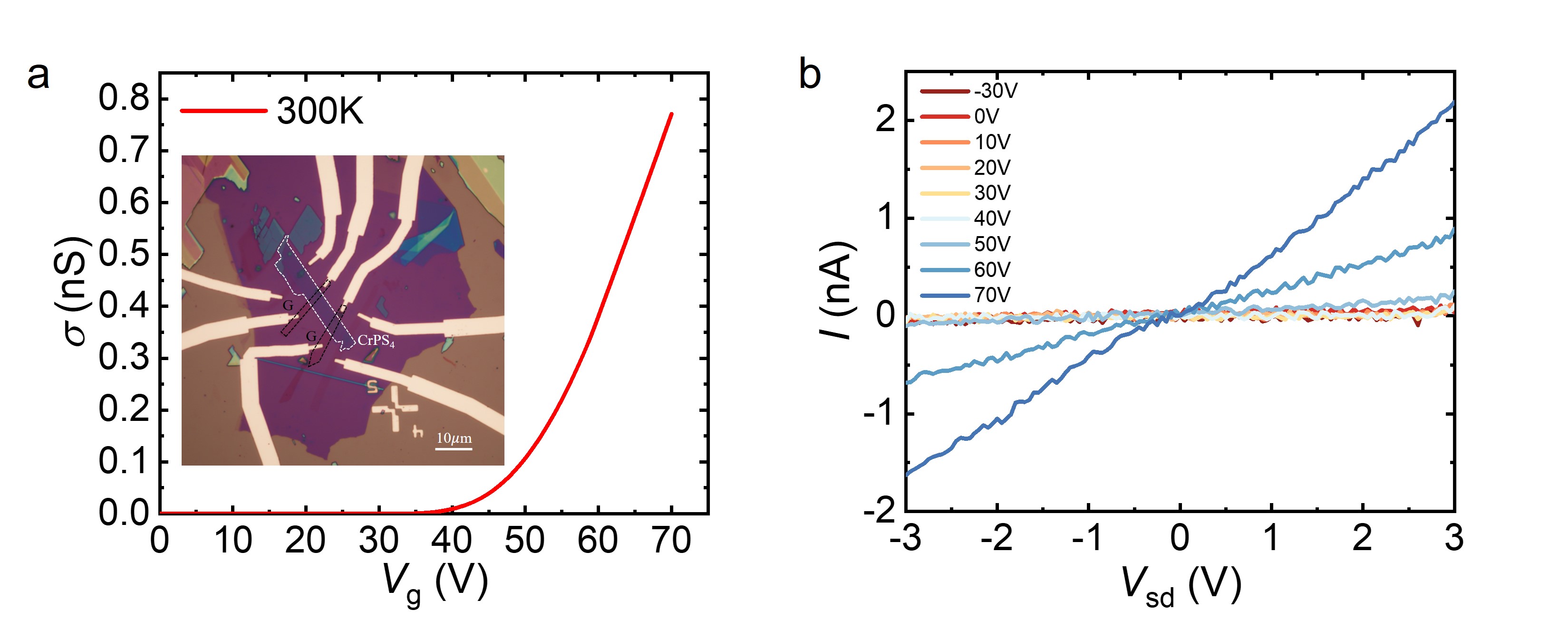}
\caption{\textbf{Field effect transistor (FET) with CrPS$_4$}. a. Gate voltage dependence of conductance measured at 300 K with bias of 3 V. The inset show the optical image of FET. The CrPS$_4$ flake is insulating when the gate voltage is zero. b. Voltage dependence of current at different back gate voltages.} 
\end{figure}

\clearpage
\newpage
\section{Additional electronic transport data of vertical junctions}

In this section we present additional electronic transport data of vertical junctions that not include in main text. Supplementary Figure 5 to Figure 12 show the data of 18, 13, 12, 10, 6, 3 and 2 layers, respectively. In all devices, there is a feature at around Néel temperature in temperature dependence of resistance measurement, and MR oscillations are observed. In addition, Supplementary Fig. 13 shows the MR data of monolayer CrPS$_4$ device. Supplementary Fig. 14 shows the AC measurements of 10-layer CrPS$_4$ device, revealing that the imaginary component is negligible compared to the real part.

\begin{figure}[H]
\centering
\includegraphics[width =1\linewidth]{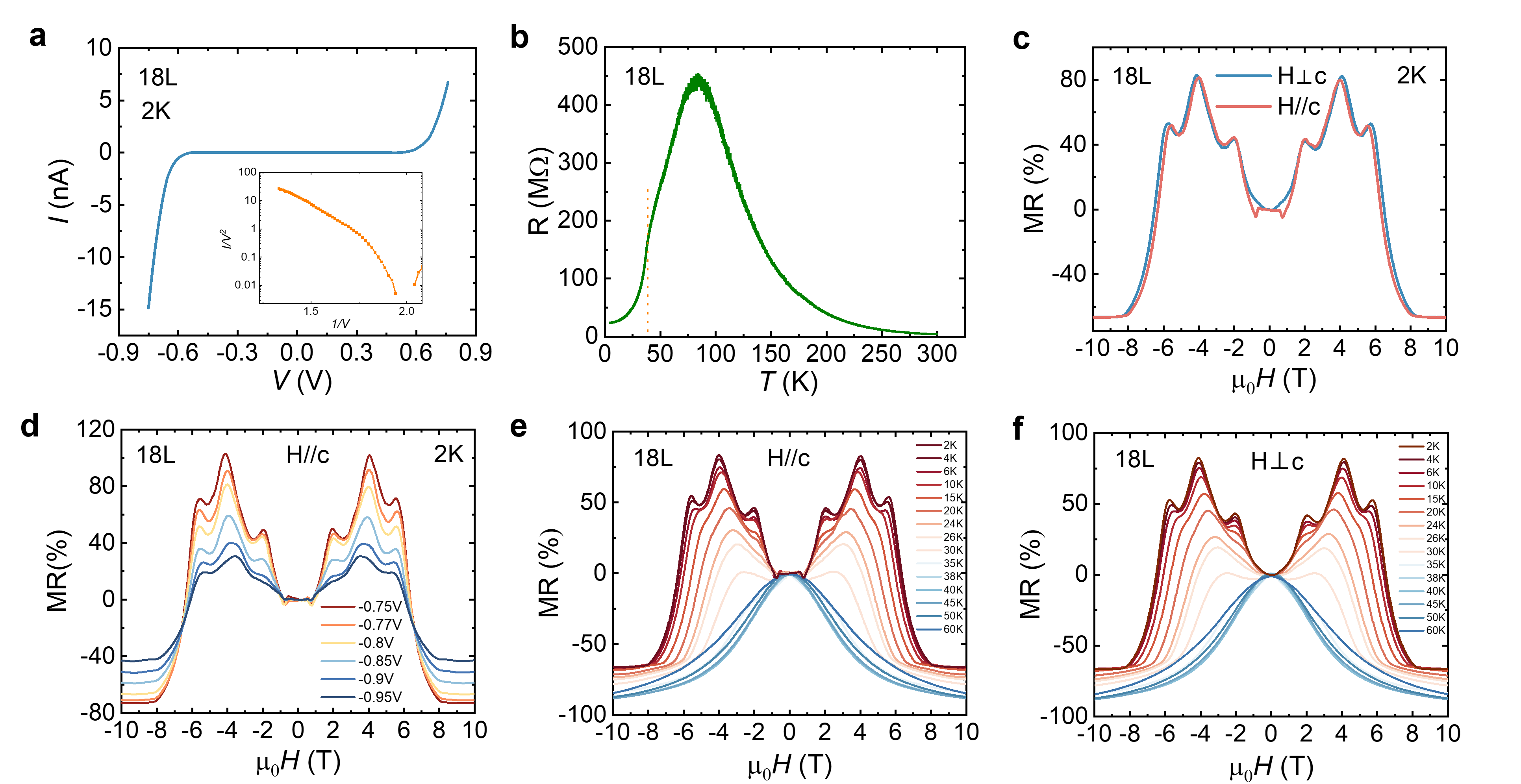}
\caption{\textbf{Transport data of 18 layers device}. a. $I-V$ curve measured at 2 K. The inset shows the plot in terms of $ln(I/V^2)$ versus $1/V$, characteristic of Fowler-Nordheim (FN) tunneling at high voltage. b. Temperature dependence of resistance at zero field. Resistance begins to decrease at around 80 K, which is not clearly understood yet. However, a signature at Néel temperature (dashed line) is still observed. c. MR at 2 K with magnetic field perpendicular and parallel to c-axis. d. MR at 2 K with different bias voltages. e. MR at different temperature when magnetic field is parallel to c-axis. f. MR at different temperature when magnetic field is perpendicular to c-axis.} 
\end{figure}

\begin{figure}[H]
\centering
\includegraphics[width =1\linewidth]{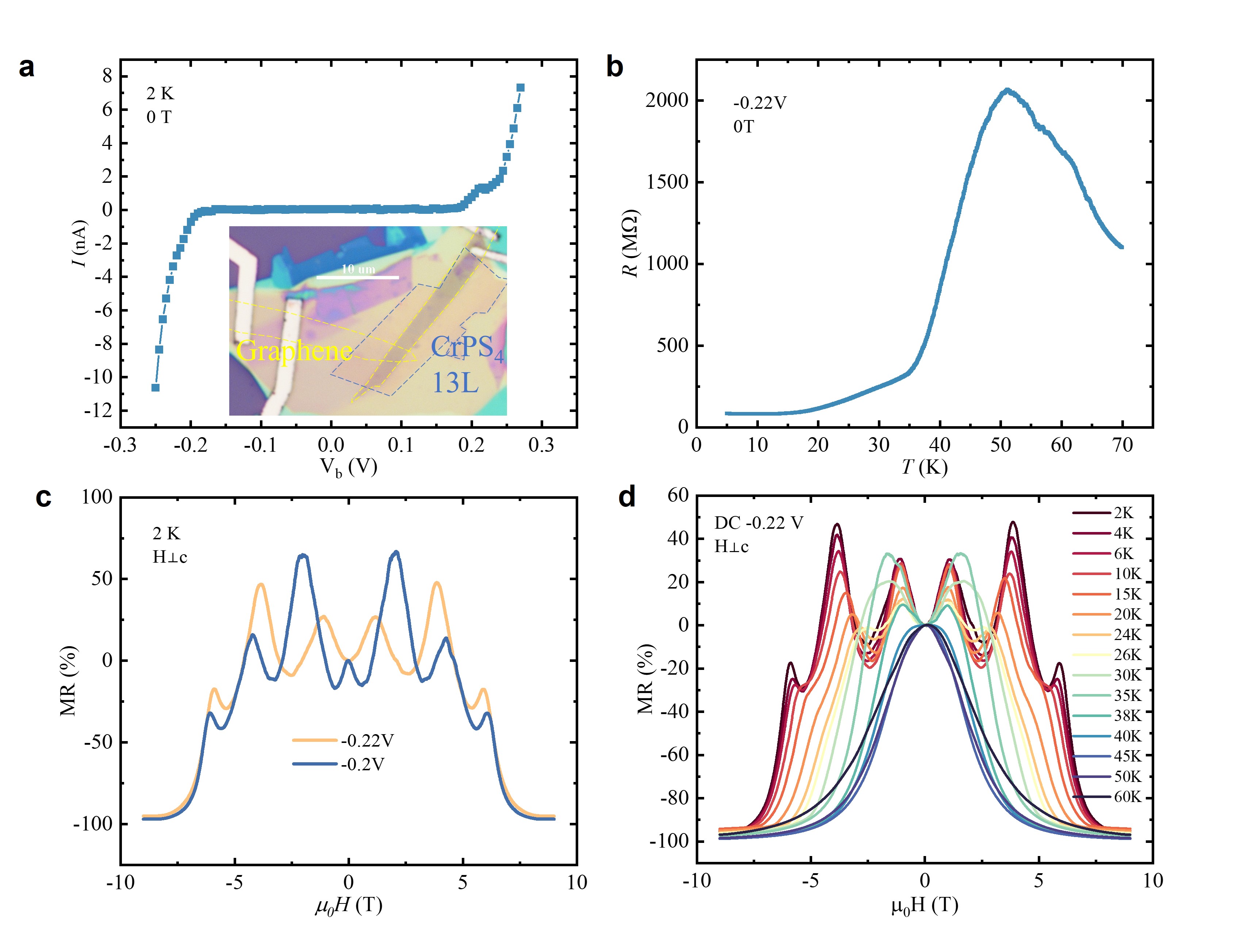}
\caption{\textbf{Transport data of 13 layers device}. a. $I-V$ curve measured at 2 K. The inset shows the optical image of this device. b. Temperature dependence of resistance at zero field. c. MR at 2 K with magnetic field perpendicular and parallel to c-axis, measured with bias voltage of -0.2 V and -0.22 V. d. MR at different temperature when magnetic field is parallel to c-axis.} 
\end{figure}

\begin{figure}[H]
\centering
\includegraphics[width =1\linewidth]{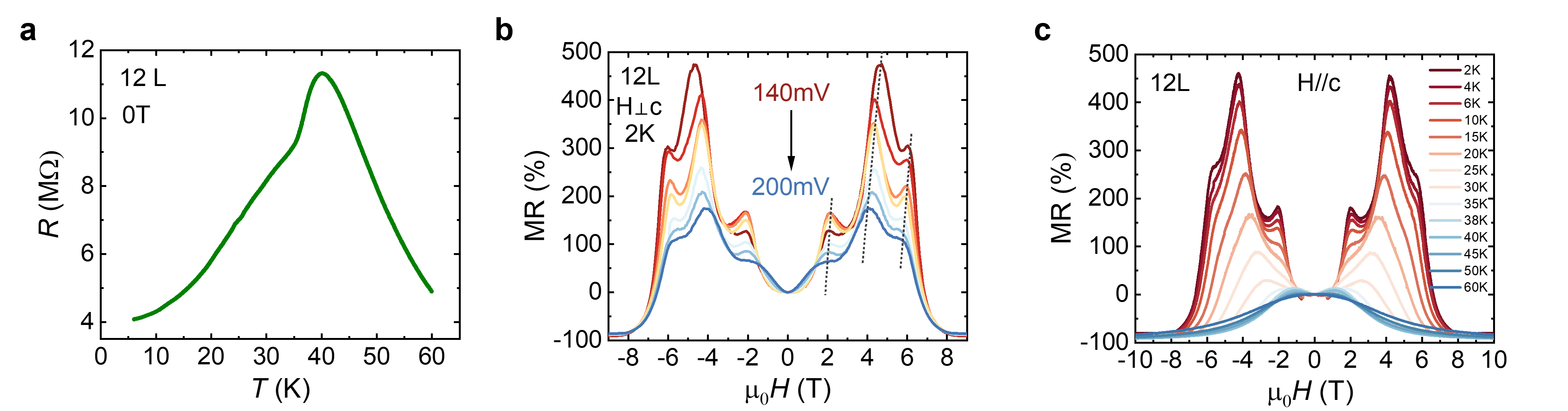}
\caption{\textbf{Transport data of 12 layers device}. a. Temperature dependence of resistance at zero field. b. MR at 2 K with different bias voltages. c. MR at different temperature when magnetic field is parallel to c-axis.} 
\end{figure}

\begin{figure}[H]
\centering
\includegraphics[width =0.9\linewidth]{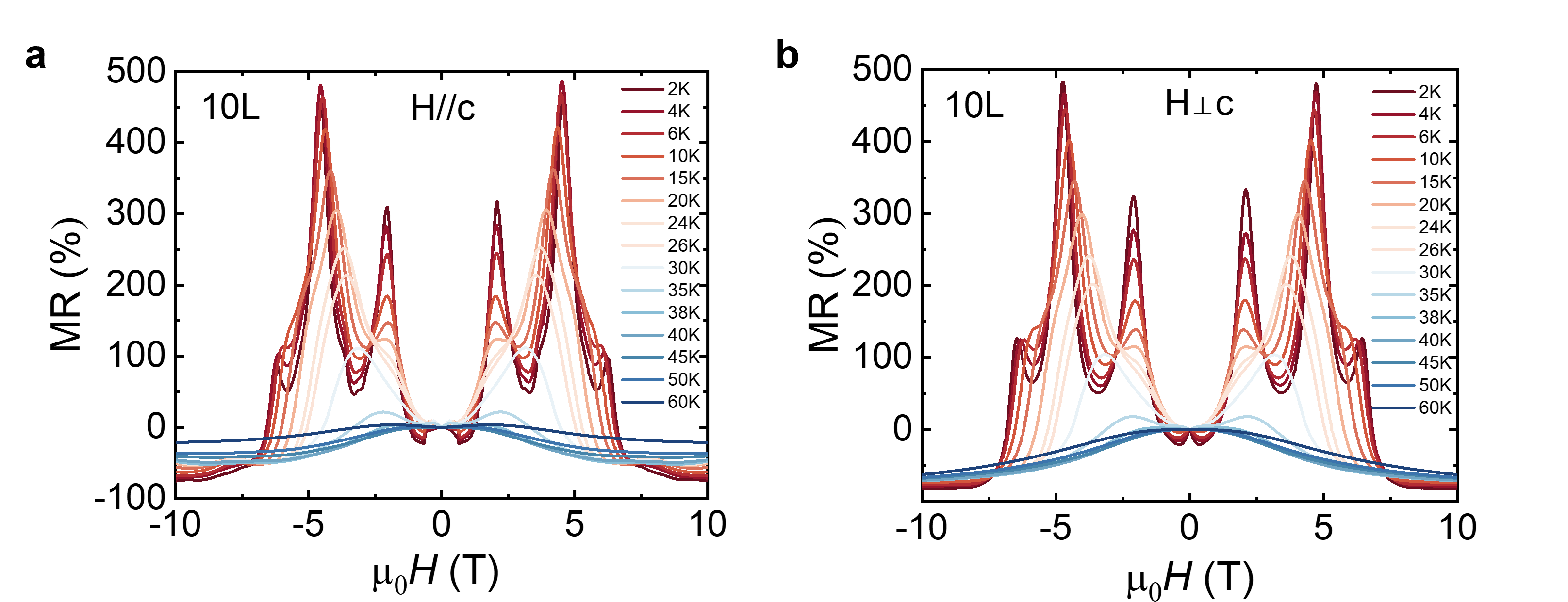}
\caption{\textbf{Transport data of 10 layers device}. a. MR at different temperature when magnetic field is parallel to c-axis. b. MR at different temperature when magnetic field is perpendicular to c-axis.} 
\end{figure}

\begin{figure}[H]
\centering
\includegraphics[width =1\linewidth]{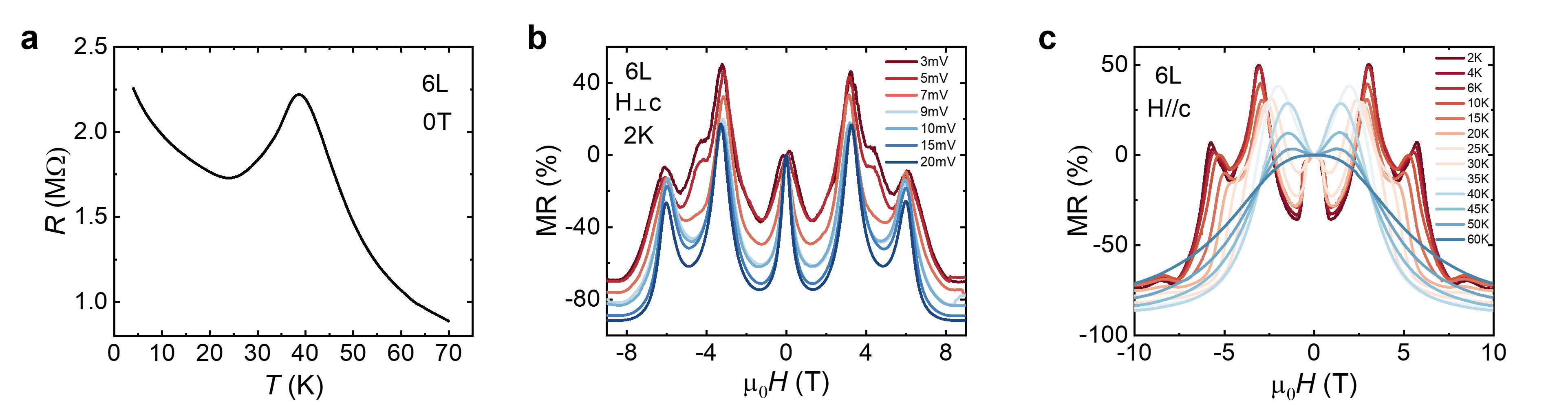}
\caption{\textbf{Transport data of 6 layers device}. a. Temperature dependence of resistance at zero field. b. MR at 2 K with different bias voltages. c. MR at different temperature when magnetic field is parallel to c-axis.} 
\end{figure}

\begin{figure}[H]
\centering
\includegraphics[width =0.5\linewidth]{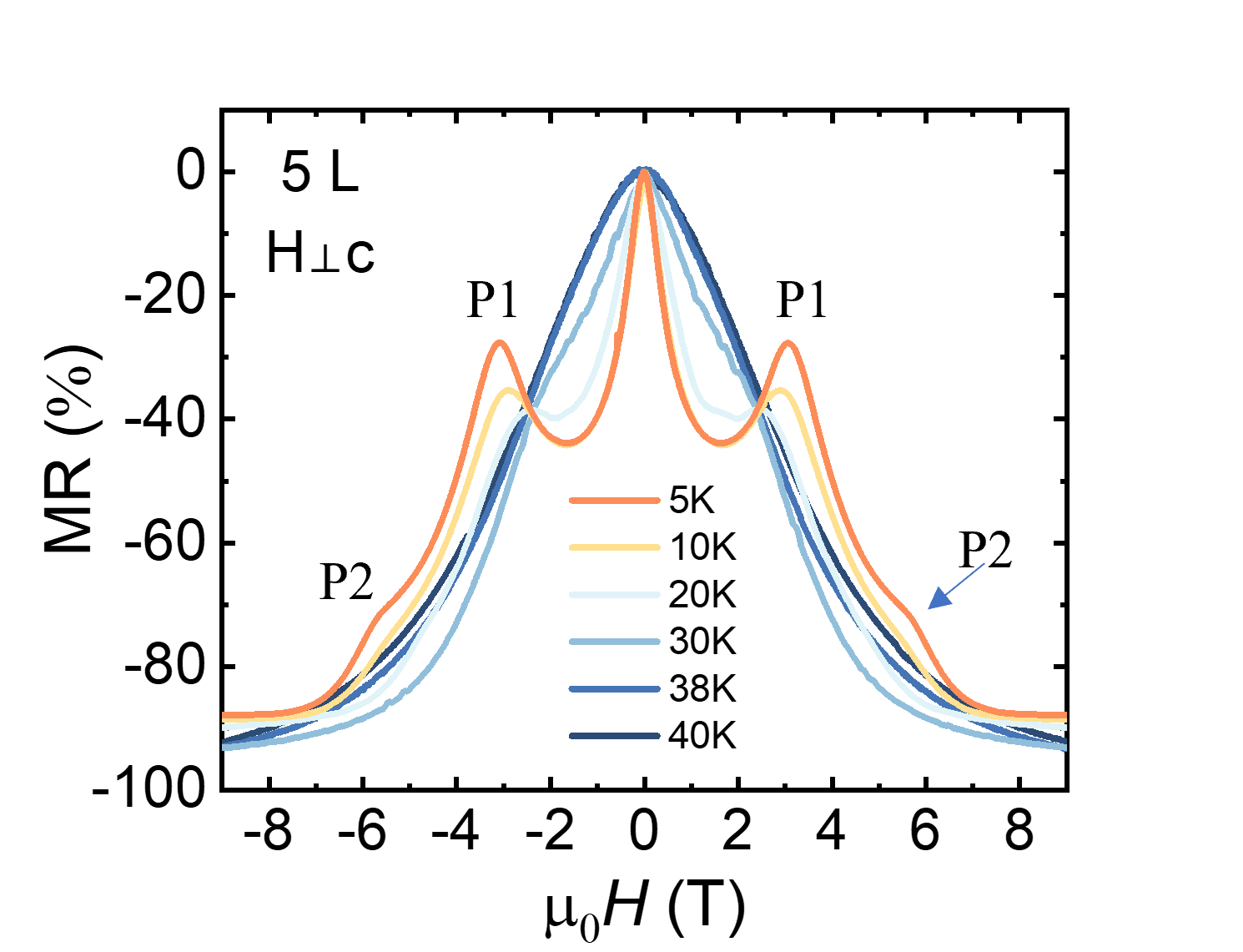}
\caption{\textbf{Transport data of 5 layers device}. MR measurements were conducted at various temperatures with the magnetic field perpendicular to the c-axis. In the resulting data, two distinct MR peaks are observed and denoted as P1 and P2 in the figure. Notably, P2 exhibits a broad peak shape, and its position can be accurately determined through differential analysis. } 
\end{figure}

\begin{figure}[H]
\centering
\includegraphics[width =1\linewidth]{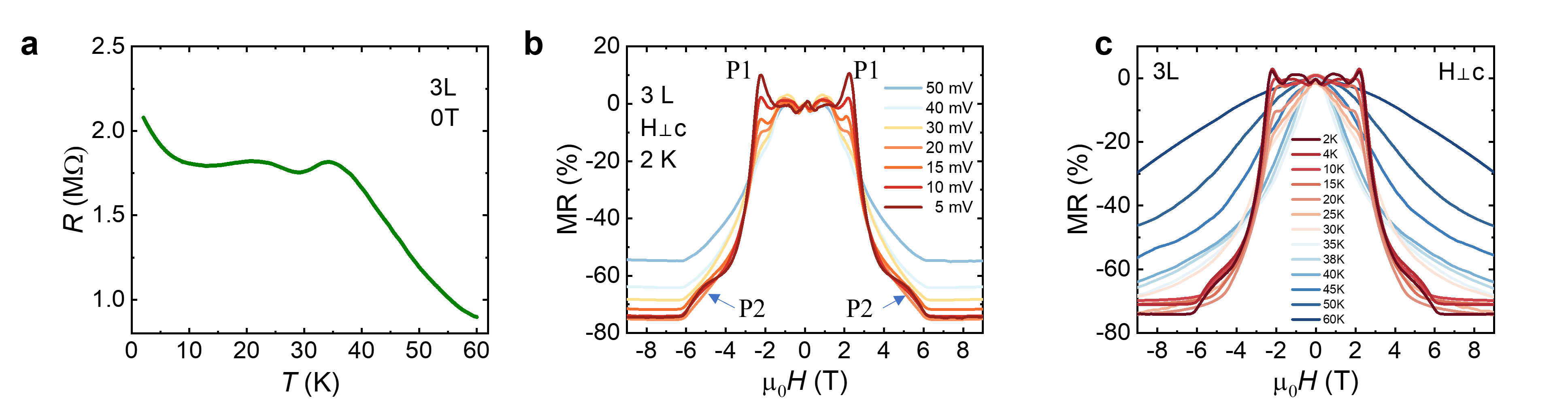}
\caption{\textbf{Transport data of 3 layers device}. a. Temperature dependence of resistance at zero field. b. MR at 2 K measured with different bias voltages when magnetic field is perpendicular to c-axis. The behavior is similar as 5 layers device, that is, P2 exhibits a broad shape. c. MR at different temperature with bias of 10 mV.} 
\end{figure}

\begin{figure}[H]
\centering
\includegraphics[width =0.8\linewidth]{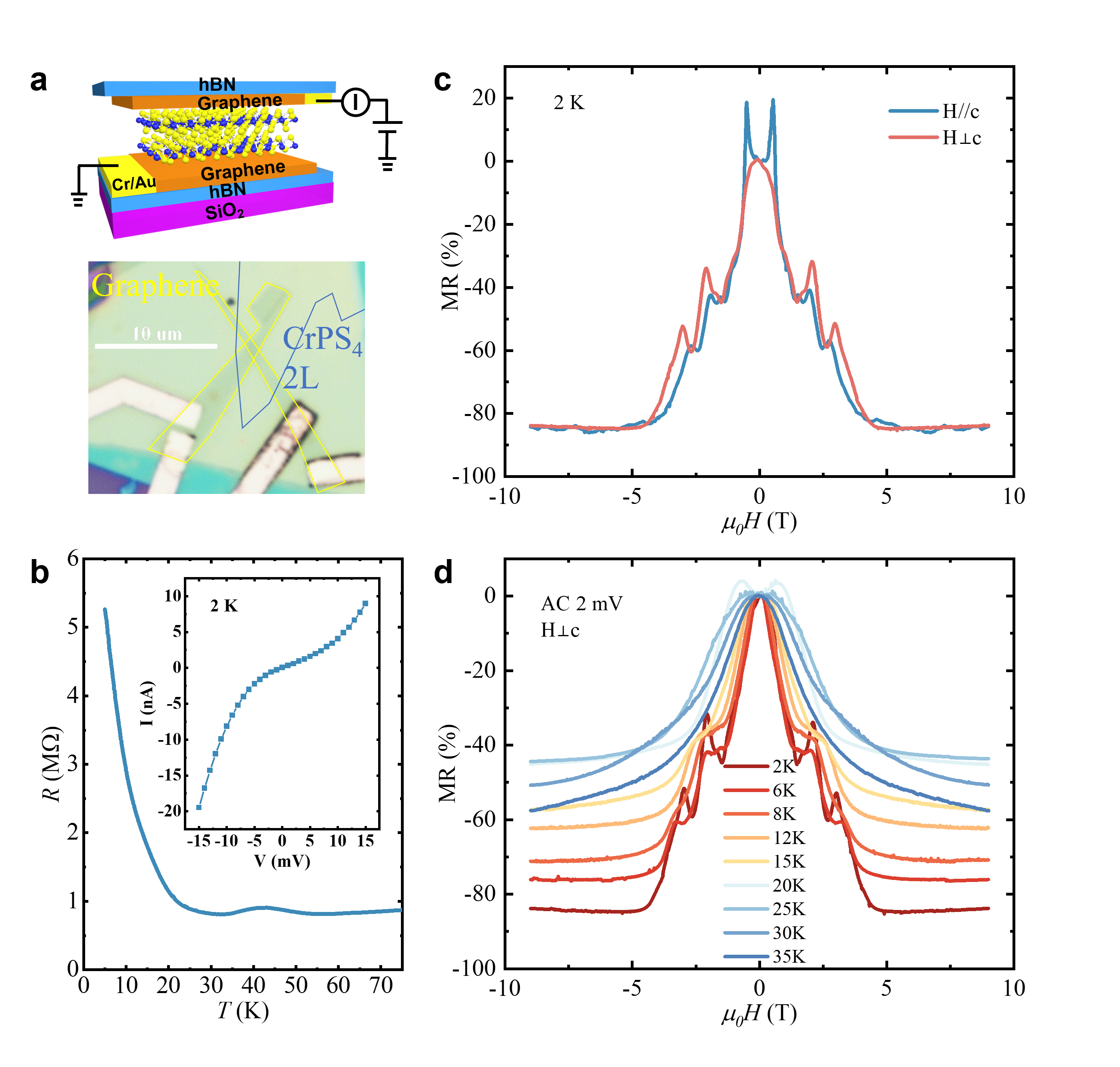}
\caption{\textbf{Transport data of 2 layers device}. a. Schematic and optical image of the device. b. Temperature dependence of resistance at zero field. The inset show the IV curve at 2 K. c. MR at 2 K measured with magnetic field parallel and perpendicular to c-axis. MR peaks are observed in both case. c. MR at different temperature with bias of 2 mV.} 
\end{figure}

\begin{figure}[H]
\centering
\includegraphics[width =0.5\linewidth]{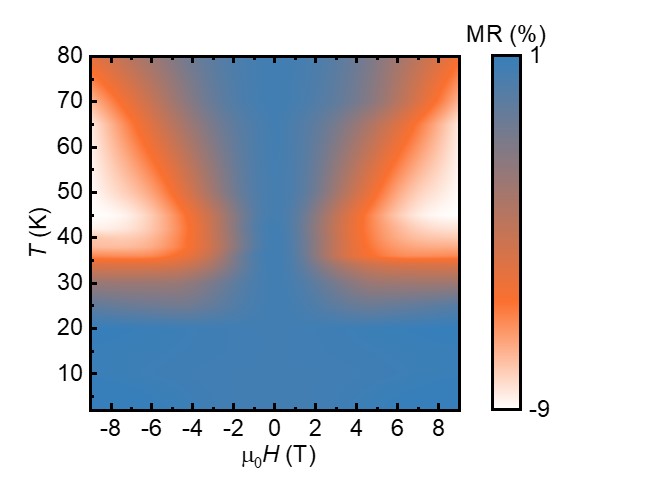}
\caption{\textbf{MR of mono-layer device}. MR at different temperature plot in 2D map. MR is predominantly negative, with the absolute value of MR peaking around the Curie temperature, forming a characteristic triangular shape. This behavior is very similar to vertical junctions with other 2D ferromagnetic semiconductor, such as CrBr$_3$ and mono-layer CrSBr.} 
\end{figure}

\begin{figure}[H]
\centering
\includegraphics[width =1\linewidth]{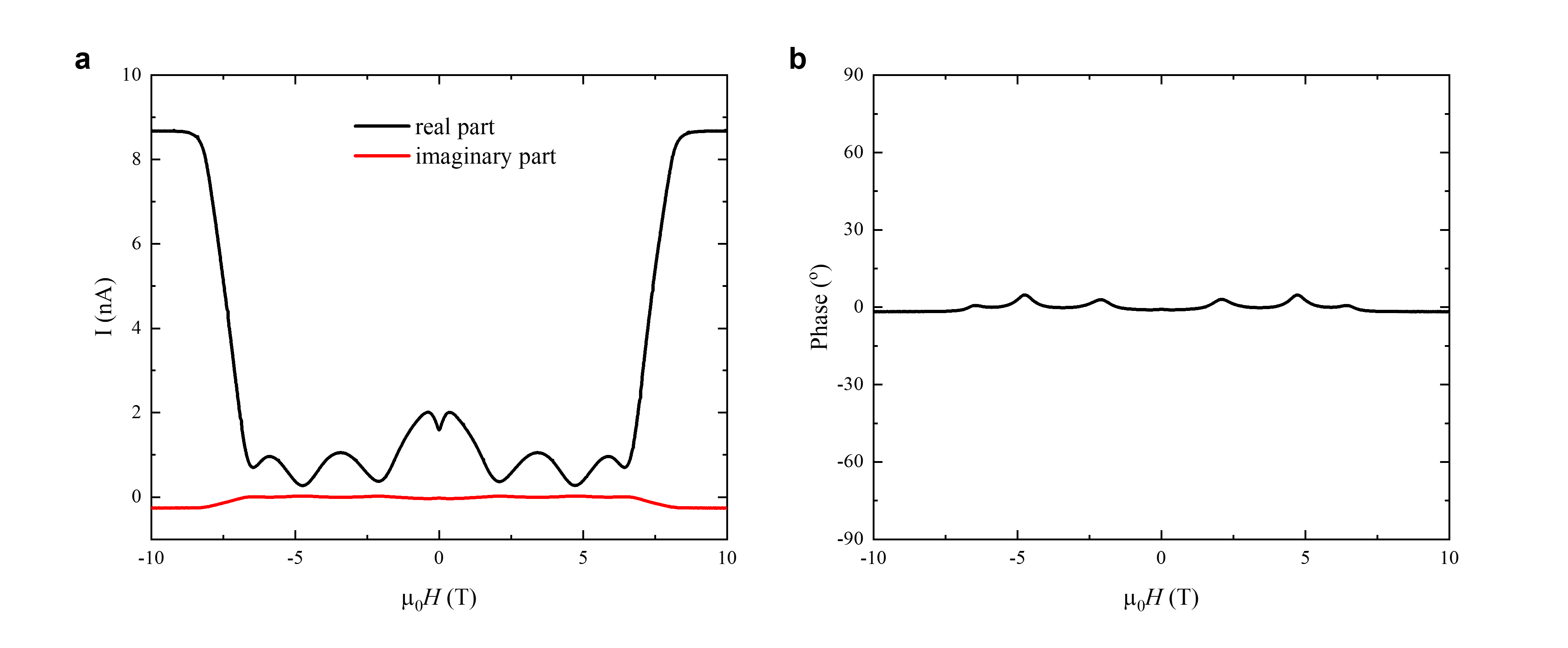}
\caption{\textbf{AC measurement of 10 L device}. a. Magnetic field dependence of the real and imaginary part of measured current. The data is from the 10 layer device at 2 K and magnetic field perpendicular to c-axis. b. Magnetic field dependence of the lock-in phase.} 
\end{figure}

\clearpage
\newpage

\section{Layer number dependence of spin-flip field in CrPS$_4$}

Considering the fact that the intralayer exchange coupling is much larger than the interlayer and anisotropic energy, each layer of CrPS$_4$ can be treated as a macro-spin. Hence we exploit the antiferromagnetic linear chain model to investigate the magnetization configuration in CrPS$_4$ ~\cite{CrCl3NN, PRXyeyu}. We consider the magnetic field applied in-plane (${\perp}c$), the energy can be written as,
\begin{equation}
E = J{S^2}\sum\limits_{i = 1}^{n - 1} {{\rm{cos}}\left( {{\theta _i} - {\theta _{i + 1}}} \right)}  + K{S^2}\sum\limits_{i = 1}^n {{\rm{co}}{{\rm{s}}^2}\left( {{\theta _i}} \right)}  - Bg{\mu _B}S\sum\limits_{i = 1}^n {{\rm{cos}}{\theta _i}} ,
\end{equation}
where ${\theta}$ is defined as the angle between spin orientation and magnetic field; first term at the r.h.s. is the antiferromagnetic coupling term, the second one is the anisotropic energy and the last one is the Zeeman energy. By defining $j = JS/g{\mu}_B/$1T and $k=KS/g{\mu}_B/$1T and $b =B$/1T, we have the dimensionless energy $\widetilde E$,
\begin{equation} \label{energy}
\widetilde E = j\sum\limits_{i = 1}^{n - 1} {{\rm{cos}}\left( {{\theta _i} - {\theta _{i + 1}}} \right)}  + k\sum\limits_{i = 1}^n {{\rm{co}}{{\rm{s}}^2}\left( {{\theta _i}} \right)}  - b\sum\limits_{i = 1}^n {{\rm{cos}}{\theta _i}}.
\end{equation}
For bulk sample, the in-plane spin flip field is given by $4j+2k$ and out-of-plane spin flip field is $4j-2k$ and spin flop field is $2\sqrt {k(k + j)}$. Based on the bulk magnetization measurement results, we have $j = 2.08$, and $k = 0.059$, which translate to $J = 0.16$ meV and $K=0.0045$ meV, very close to the neutron scattering measurement results ~\cite{CPSPRB2020}.

To determine the layer dependency of spin-flip field of CrPS$_4$, we follow the steps in Ref ~\cite{CrCl3NN} and calculate the $M$ matrix with matrix element defined by $M_{ij} = {\partial ^2}\widetilde E/\partial {\theta _i}\partial {\theta _j}$ and all the spins pointed parallel with magnetic field, i.e., ${\theta}_i = 0$,
\begin{equation}
M = \left( {\begin{array}{*{20}{c}}
{ - j + b + 2k}&j&0&0&0\\
j&{ - 2j + b + 2k}&j&0&0\\
0& \ddots & \ddots & \ddots &0\\
0&0&j&{ - 2j + b + 2k}&j\\
0&0&0&j&{ - j + b + 2k}
\end{array}} \right)
\end{equation}
which has to be positive definite for field $b >$ the critical spin-slip field $b_{\rm{flip}}$. Setting smallest eigenvalue equals to 0 gives the critical field,
\begin{equation}
{b_{{\rm{flip}}}} = 2j\left( {1 + \cos \frac{\pi }{n}} \right) + 2k
\end{equation}
This layer dependent critical field corresponds to the theoretical curve plotted in Figure 3d in main text.

\clearpage
\newpage
\section{Calculated magnetic field dependence of magnetization configuration and tunneling conductance}

Based on the determined $j= 2.08$ and $k = 0.059$, we can simulate the field dependency of magnetization of CrPS$_4$ with specific layer numbers. The solution of $\theta_i$ can be obtained by minimizing the energy of Eq. (\ref{energy}), we have simulated cases with 6-layer and 10-layer CrPS$_4$ as shown in Supplementary Figure 15a, where the label denotes the layer number. We notice that the $\theta_i$ is not monotonically changed by the magnetic field.

\begin{figure}[H]
\centering
 \includegraphics[width =0.9\linewidth]{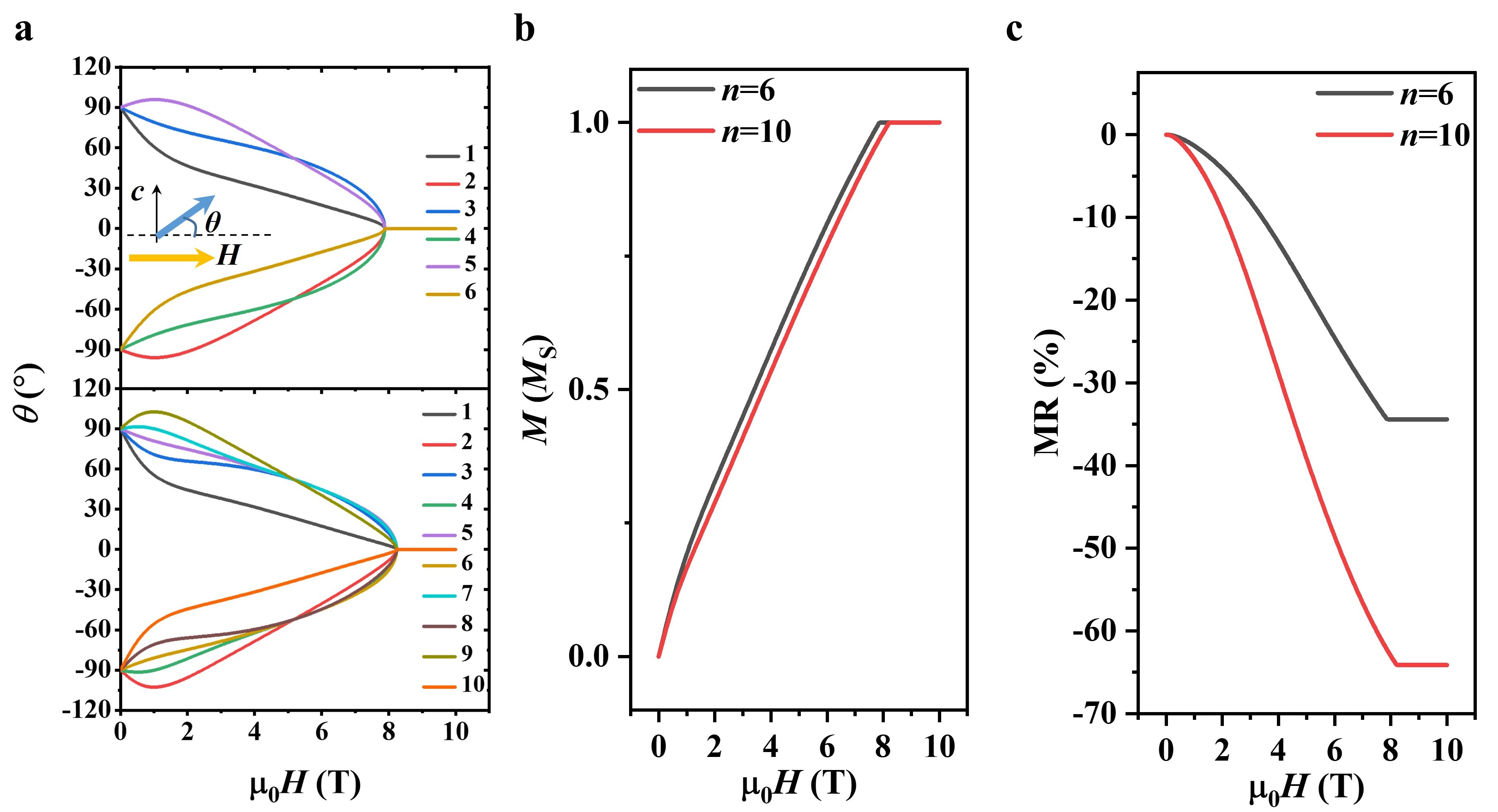}
 \caption{\textbf{Simulation results of 6 and 10 layers CrPS$_4$}. a Simulated $\theta_i$ as a function of magnetic field for $n$ =6 (upper panel) and $n$ = 10 (down panel) where labels on the right denote the layer number. b Simulated magnetization as a function of external magnetic field for 6 and 10 layers CrPS$_4$. c MR calculated as a function of magnetic field based on tunneling scheme for 6 and 10 layers CrPS$_4$.} 
\end{figure}

Now we derive the magneto-conductance based on the tunneling scheme, that is, the electrons tunneling through spin-split band of CrPS$_4$. We consider a bias voltage $V$, the barrier would be in trapezoid shape,
\begin{equation}
U(x) = \begin{cases}
U = 0, & x \le 0 \\
U = {\phi _0} - eV \cdot x/d, & 0 < x \le d \\
U =  - eV,& x > d \\
\end{cases}
\end{equation}
where ${\phi}_0$ is the barrier height when $V = 0$ and $d$ is the thickness of barrier layers, the eigen wavefunction should be the linear combination of Airy functions $\psi  = a{\rm{Ai}}\left( z \right) + b{\rm{Bi}}\left( z \right)$, and the variable $z$ is defined as,

\begin{equation}
z =  - \lambda \left( {x - \frac{{{\phi _0} - E}}{{eV}}d} \right){\rm{  with  }} \lambda  = {\left( {\frac{{2meV}}{{{\hbar ^2}d}}} \right)^{1/3}}
\end{equation}

We consider the band splitting for parallel and antiparallel case,${\phi _0} \pm \Delta$ , meaning CrPS$_4$ are at antiferromagnetic and ferromagnetic state respectively. By exploiting the transfer matrix technique, we obtain the transfer matrix between $m$th layer and ($m$+1)th layer to be,
\begin{equation}
P = \pi \sin \frac{\theta }{2}\left( {\begin{array}{*{20}{c}}
{1/\pi /\tan \frac{\theta }{2}}&{{\rm{A}}{{\rm{i}}_ + }{\rm{Bi}}{{\rm{'}}_ - } - {\rm{Ai}}{{\rm{'}}_ + }{\rm{B}}{{\rm{i}}_ - }}&0&{{\rm{B}}{{\rm{i}}_ + }{\rm{Bi}}{{\rm{'}}_ - } - {\rm{Bi}}{{\rm{'}}_ + }{\rm{B}}{{\rm{i}}_ - }}\\
{{\rm{Ai}}{{\rm{'}}_ - }{\rm{B}}{{\rm{i}}_ + } - {\rm{A}}{{\rm{i}}_ - }{\rm{Bi}}{{\rm{'}}_ + }}&{1/\pi /\tan \frac{\theta }{2}}&{{\rm{Ai}}{{\rm{'}}_ - }{\rm{B}}{{\rm{i}}_ + } - {\rm{A}}{{\rm{i}}_ - }{\rm{Bi}}{{\rm{'}}_ + }}&0\\
0&{{\rm{Ai}}{{\rm{'}}_ + }{\rm{A}}{{\rm{i}}_ - } - {\rm{A}}{{\rm{i}}_ + }{\rm{Ai}}{{\rm{'}}_ - }}&{1/\pi /\tan \frac{\theta }{2}}&{{\rm{Bi}}{{\rm{'}}_ + }{\rm{A}}{{\rm{i}}_ - } - {\rm{B}}{{\rm{i}}_ + }{\rm{Ai}}{{\rm{'}}_ - }}\\
{{\rm{Ai}}{{\rm{'}}_ + }{\rm{A}}{{\rm{i}}_ - } - {\rm{A}}{{\rm{i}}_ + }{\rm{Ai}}{{\rm{'}}_ - }}&0&{{\rm{Ai}}{{\rm{'}}_ + }{\rm{B}}{{\rm{i}}_ - } - {\rm{A}}{{\rm{i}}_ + }{\rm{Bi}}{{\rm{'}}_ - }}&{1/\pi /\tan \frac{\theta }{2}}
\end{array}} \right)
\end{equation}
where $\theta$ is the angle between two layers’ spin polarizations, the subset $\pm$ indicates the variable of the corresponding Airy function is 
${z_ \pm } =  - \lambda \left( {m{d_1} - \frac{{{\phi _0} \pm \Delta  - {E_0}}}{{eV}}n{d_1}} \right)$  with $d_1$ denoting the thickness of each CrPS$_4$ layer; the prime denotes the first derivative of Airy function. The transfer matrix between electrode and the outmost CrPS$_4$ layer is given by,
\begin{equation}
D = \left( {\begin{array}{*{20}{c}}
{{\rm{Ai}}\left( {{z_ - }} \right) - \lambda {\rm{Ai'}}\left( {{z_ - }} \right)/k}&0&{{\rm{Bi}}\left( {{z_ - }} \right) - \lambda {\rm{Bi'}}\left( {{z_ - }} \right)/k}&0\\
0&{{\rm{Ai}}\left( {{z_ + }} \right) - \lambda {\rm{Ai'}}\left( {{z_ + }} \right)/k}&0&{{\rm{Bi}}\left( {{z_ + }} \right) - \lambda {\rm{Bi'}}\left( {{z_ + }} \right)/k}\\
{{\rm{Ai}}\left( {{z_ - }} \right) + \lambda {\rm{Ai'}}\left( {{z_ - }} \right)/k}&0&{{\rm{Bi}}\left( {{z_ - }} \right) + \lambda {\rm{Bi'}}\left( {{z_ - }} \right)/k}&0\\
0&{{\rm{Ai}}\left( {{z_ + }} \right) + \lambda {\rm{Ai'}}\left( {{z_ + }} \right)/k}&0&{{\rm{Bi}}\left( {{z_ + }} \right) + \lambda {\rm{Bi'}}\left( {{z_ + }} \right)/k}
\end{array}} \right) \cdot \sqrt k /2
\end{equation}
where ${\rm{ }}{z_ \pm } = \lambda \frac{{{\phi _0} \pm \Delta  - {E_0}}}{{eV}}d$ for left electrode to 1st layer CrPS$_4$ and for right electrode to $n$th layer electrode is ${\rm{ }}{z_ \pm } = \lambda n{d_1}\left( {\frac{{{\phi _0} \pm \Delta  - {E_0}}}{{eV}} - 1} \right)$  . The overall transfer matrix $T$ is a 4 by 4 matrix can be expressed as,
\begin{equation}
T = D\prod\limits_{i = 1}^{n - 1} {P\left( {{\theta _i} - {\theta _{i + 1}}} \right)} {D^{ - 1}}.
\end{equation}

The overall matrix $T$ satisfies,
\begin{equation}
\left( \begin{array}{l}
{A_{L \uparrow }}\\
{A_{L \downarrow }}\\
{A_{R \uparrow }}\\
{A_{R \downarrow }}
\end{array} \right) = T\left( \begin{array}{l}
{t_ \uparrow }\\
{t_ \downarrow }\\
0\\
0
\end{array} \right) = \left( {\begin{array}{*{20}{c}}
{{T_{11}}}&{{T_{12}}}\\
{{T_{21}}}&{{T_{22}}}
\end{array}} \right)\left( \begin{array}{l}
{t_ \uparrow }\\
{t_ \downarrow }\\
0\\
0
\end{array} \right)
\end{equation}

where ${A_{L \uparrow }}$ and ${A_{R \uparrow }}$  denotes the spin up incident and reflected wave coefficients. If we set the incident wave to be ${\left( {\begin{array}{*{20}{c}}
{{A_{L \uparrow }}}&{{A_{L \downarrow }}}
\end{array}} \right)^T} = {\left( {\begin{array}{*{20}{c}}
{\cos \varphi }&{\sin \varphi }
\end{array}} \right)^T}$, the overall conductance is given by,
\begin{equation}
G = \frac{1}{{2\pi }}\int_0^{2\pi } {\left( {{{\left| {{t_ \uparrow }} \right|}^2} + {{\left| {{t_ \downarrow }} \right|}^2}} \right)d\varphi }  = \frac{1}{2}{\rm{Tr}}\left( {T_{11}^{ - 1}} \right)
\end{equation}

We choose the effective mass of CrPS$_4$ to be $2m_e$, Fermi energy of electrodes to be $E_0=0.2$ eV and effective mass to be $0.039m_e$ for few layer graphene, barrier potential ${\phi}_0 = 0.5$ eV and ${\Delta} = 0.05$ eV. We should point it out that the exact value of these parameters only affect the magnitude of MR, here we are exploring MR oscillation is possible or not in the tunneling scenario. By exploiting the magnetic field dependency of spin orientation angle of each CrPS$_4$ layer, we can obtain the magneto-conductance of CrPS$_4$ with $MR = (R(B)-R(0))/R(0)$, as shown in Supplementary Figure 15(c), no MR oscillation is observed. 

\clearpage
\newpage
\section{Defect states in CrPS$_4$}
We constructed a bi-layer CrPS$_4$ system to investigate the effects of vacancies. Given that the ground state of CrPS$_4$ is known to be an A-type antiferromagnet, we initiated our calculations with antiferromagnetic ordering between each layer. Sulfur vacancies are commonly found in transition metal chalcogenides, and thus, we assume S vacancies are the main defect in CrPS$_4$. To simulate the vacancy effect, we removed one S atom from the bi-layer CrPS$_4$ system (total of 16 S atoms), as illustrated in Supplementary Fig. 16. We performed first-principles calculations using the Vienna Ab-initio Simulation Package (VASP) \cite{VASP1,VASP2}. All calculations were based on the generalized gradient approximation (GGA) in the Perdew–Burke–Ernzerhof (PBE) form \cite{PBE} and employed a plane-wave basis set within the projector augmented wave (PAW) method \cite{PAW1,PAW2}. The cut-off energy for the basis set was set to 500 eV. The convergence criterion for the electron density self-consistency cycles was 10$^{-8}$ eV and for the structural relaxation was 10$^{-2}$ eV/\AA.

\begin{figure}[H]
\centering
 \includegraphics[width =0.9\linewidth]{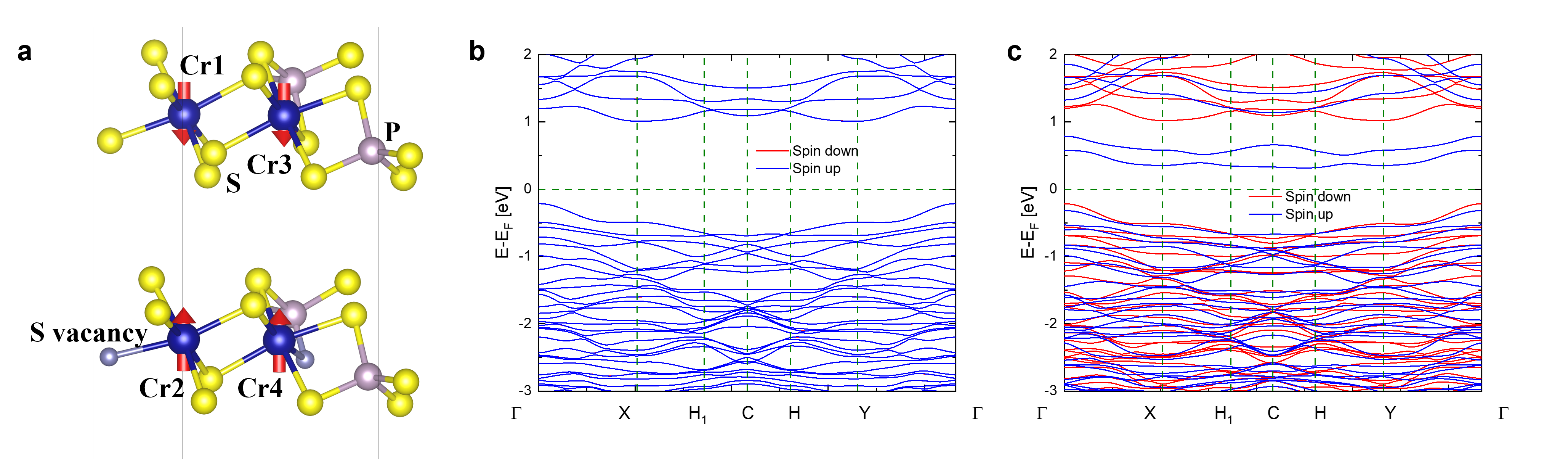}
 \caption{\textbf{Calculated band structures}. a. The crystal structure of bilayer CrPS$_4$ with one S vacancy. b. The band structure of bilayer CrPS$_4$ without defect in the antiferromagnetic state. c. The band structure of bilayer CrPS$_4$ with one S vacancy in the lower layer as illustrated in a.} 
\end{figure}

The calculated magnetization values are presented in Supplementary Table 3. The first row displays the magnetization of each layer in the absence of defects, while the second row shows the magnetization of CrPS$_4$ with a sulfur vacancy in the lower layer, specifically affecting the Cr2 and Cr4 atoms. Our findings indicate that the magnetization of the lower layer with the sulfur vacancy increased by approximately 0.1 $\mu$B/Cr, whereas the magnetization of the upper layer remained nearly unchanged.

\renewcommand{\figurename}{{\bf Supplementary Table.}}

\begin{figure}[H]
\addtocounter{figure}{-14}
\centering
 \includegraphics[width =0.7\linewidth]{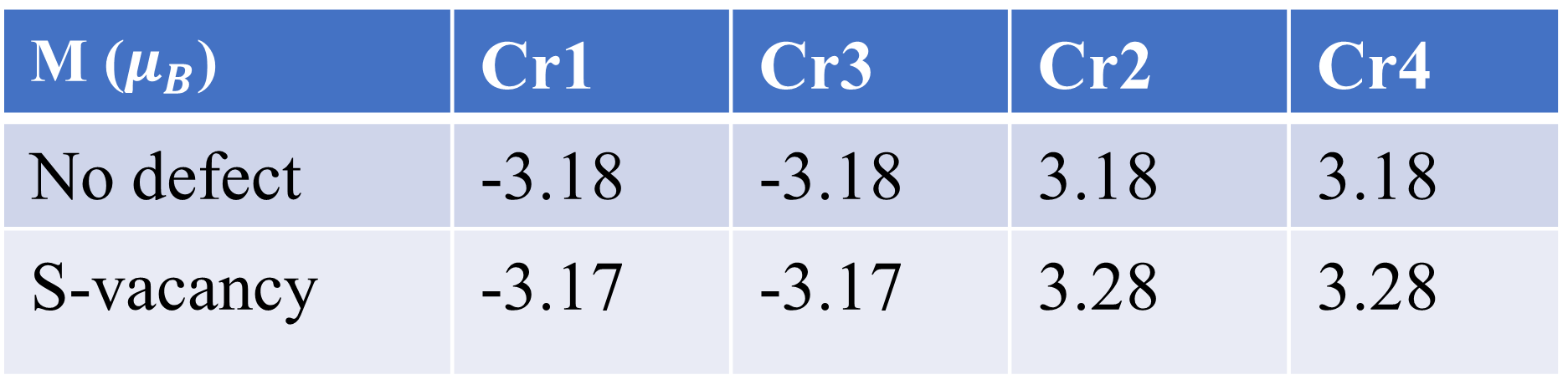}
 \caption{\textbf{Calculated magnetization of bilayer CrPS$_4$ without and with S vacancy.}. } 
\end{figure}

\renewcommand{\figurename}{{\bf Supplementary Fig.}}
The calculated band structure of bi-layer CrPS$_4$ is shown in Supplementary Fig. 16c. For the system with S-vacancy, degenerate spin states appear inside the band gap of CrPS$_4$. This defect state is fully spin polarized and in the same direction of magnetization of Cr atoms in the same layer.

\clearpage
\newpage

\section{Spin Selected Interlayer Hopping}

In this section, we propose the spin selected interlayer hopping (SSIH) model to describe the electron hopping in vertical junctions of CrPS$_4$. The SSIH proposed here contrasts with the conventional variable range hopping (VRH) model in two aspects. Firstly, the conventional VRH is built in homogeneous $d(=1,2,3)$ dimensions, here we propose SSIH in 2+1 dimensions, where 2 stands for continuous in-plane hopping component and 1 denotes the out-of-plane hopping between discrete layers. Secondly conventional VRH theory is built in spinless system and the hopping range varies at different temperatures, here the hopping range in our system is also affected by the spin-configuration as discussed in the main text. So two effects of magnetic field are considered. a) On one side, it affects the spin-configuration in CrPS$_4$, thereby resulting in different density of states (DOS).  b) On the other side, the magnetic field also contracted the wavefunction, i.e.,the localization length decreases as magnetic field increases, which is also expected in spinless system. We adopt Shklovskii and Efros's asymptotic description of wavefunction under magnetic field to model this effect. Combining these competing effects, we show that the MR oscillations and electric field dependency can be well explained by this SSIH model. \comment{Furthermore, we notice that the position of MR peaks can be more precisely described when we consider the interference between spin Berry phases\cite{spinphase}, whose origin needs further evaluation. We also show taking this phase into considerations to interpret our data at the end of this section}.

Before delving into SSIH model, we first briefly discuss the conventional VRH of localized states. The conductance $G\propto \exp(-\xi)$ is determined by the dimensionless hopping range $\xi$, which is defined as,
\begin{equation}
\xi = {r  \over a} + {{\epsilon}  \over {k_{\rm B}T}},
\end{equation}
where first term describes the spatial tunneling with $a$ to be the localization length, and second term denotes thermal activation with $k_{\rm B}$ to be Boltzmann constant. The hopping distance $r$ and energy $\epsilon$ are connected by the density of localized states $g$. Considering the two dimensional(2D) case, we have
\begin{equation}
r \propto 1 / N^{1/2} \propto 1 / {\left( {g\varepsilon } \right)^{1/2}}.
\end{equation}
By substituting Eq. (13) into Eq. (12) and minimize $\xi$, we obtain
\begin{equation}
\xi = \left({T_0/T}\right)^{1/3} \text{    and    } T_0=\beta/k_{\rm B}ga^2.
\end{equation}
The value of numerical coefficient $\beta$ can be obtained by using percolation approaches. The optimized hopping distance is given by 
\begin{equation}
r \propto \left({g\epsilon}\right)^{-1/2} \propto a \left(T_0/ T\right)^{1/3}.
\end{equation}
The above derivations give the temperature dependence of conductance and hopping distance $r$. 

The density of states $g$, which is treated as a constant in conventional VRH, is however both temperature and magnetic field dependent in vertical hopping conduction in CrPS$_4$. The localized states might come from S vacancy as calculated by first-principle calculations in Supplementary Note 5. The energy of the defect state in $i$th layer is correlated with the spin direction. We can set the defect energy to be $E_\uparrow$ for parallel spin and $E_\downarrow$ for anti-parallel spin ($E_\downarrow > E_\uparrow$). We consider the hopping from the localized state $\left|  \uparrow  \right\rangle$ in $i$th layer (with energy $E_\uparrow$) to the $j$th layer, and assume the magnetization angle between these two layers is $\theta$. The eigenstate in $i$th layer should be projected onto eigenstates in $j$th layer as $\cos {\theta  \over 2}\left|  \uparrow  \right\rangle  + \sin {\theta  \over 2}\left|  \downarrow  \right\rangle $, the energy would be,
\begin{equation}
{E_j} = {\cos ^2}{\theta  \over 2}{E_ \uparrow } + {\sin ^2}{\theta  \over 2}{E_ \downarrow } = {E_ \uparrow } + {\sin ^2}{\theta  \over 2}\Delta E,
\end{equation}
where $\Delta E = E_\downarrow - E_\uparrow $. Such hopping requires an increase in energy, which would be difficult for thermal excitation at low temperature. To make such hopping accessible, We apply the Gaussian disorder model (GDM), which is widely used in VRH simulations ~\cite{GDM1,GDM2}. The GDM assumes that the energy of defect states obeys the Gaussian distribution, and we therefore can assume the density of (defect) states (DOS) near Fermi energy is given by,
\begin{equation}
g =g_0 \exp\left({\left({1-\sin ^4{\theta  \over 2}}\right) {{\Delta E^2} \over {2\sigma^2}}}\right)=g_0 f(H).
\end{equation}
We can see the DOS depends on spin configurations and $g$ is a function of both temperature and magnetic field $H$. Therefore, $T_0$ in Eq. (14) is not a constant and we do not expect to see typical $\ln(\sigma) \propto T^{-1/3}$ behavior. 

As we focus on the MR effect, we consider generally two magnetic field effects. Firstly is the spin configurations due to magnetic field. As shown in Supplementary Fig. 15a, the magnetization direction of each CrPS$_4$ layer can be effectively tuned by external magnetic field, the function $f(H)$ in Eq. (17) can be numerically evaluated. At zero magnetic field, the nearest layers have opposite spin directions and hence the DOS is small, inter-next-nearest-layer hopping is preferred. As magnetic field turned on, the nearest layer's magnetization tends to align in the same direction, leading to a larger DOS and eventually inter-nearest-layer hopping dominates. 

The second effect is the contraction of localized wavefunction due to magnetic field and this shall lead to positive MR. The evolution of wavefunction under magnetic field is challenging to solve accurately, here we adopt the asymptotic behavior of wavefunction contraction by Shklovskii and Efros~\cite{Shklovskii}, where the localization length $a$ is assumed to be smaller than the magnetic length $\lambda = \sqrt{\hbar/eB}$, 
\begin{equation}
\psi \propto \exp \left[ { - \left( {{r \over a} + {{{\rho ^2}ra} \over {24{\lambda ^4}}}} \right)} \right].
\end{equation}
Here $a$ denotes the localization length without magnetic field, the second term describes the wavefunction contraction in magnetic field with $\rho$ corresponds to the distance away from the wavefuntion center in the transverse direction (perpendicular to magnetic field). Following the derivations in Ref~\cite{Shklovskii},  $\xi$ would increase as,
\begin{equation}
\xi  = \left({{T_0} \over {T}}\right)^{1/3}+t_1 {{a^4T_0} \over {{\lambda^4T}}} = a_1/f(H)^{1/3}+a_2H^2/f(H).
\end{equation}
where ${a_1} = ({\beta/k_{\rm B}T {a^2} g_0})^{1/3}$ and ${a_2} = t_1\beta a^2e^2/k_{\rm B}Tg_0\hbar^2$. $t_1$ is numerical coefficient, and the second term describes the increase of hopping range due to wavefunction contraction. 

Based on the above derivations, we simulate the MR of multilayer CrPS$_4$. We first establish the conductance between arbitrary $i$th and $j$th layer, 
\begin{equation}
G_{ij} = \exp[-(\xi_{ij}+|i-j|d/\lambda_z)],
\end{equation}
where $|i-j|d$ denotes the layer separation and $\lambda_z$ is the localization length in $z$ direction, which is in accordance with conventional VRH theory.  Based on Kirchhoff laws, the voltage and current obeys,

\begin{equation}
\sum\limits_{j \ne i}^n {{G_{ij}}\left( {{V_j} - {V_i}} \right)}= \begin{cases}
-I, & i=1 \\
0, & 1 < i < n \\
I. & i=n \\
\end{cases}
\end{equation}
The above $n$ equations are not independent, by setting $V_n=0$, we can obtain $n-1$ independent equations,
\begin{equation}
\sum\limits_{j \ne i}^{n - 1} {{G_{ij}}{V_j}}  - {V_i}\sum\limits_{j \ne i}^{n - 1} {{G_{ij}}}  = \begin{cases}
-I, & i=1 \\
0. & 1 < i < n \\
\end{cases}
\end{equation}
After solving the $n-1$ independent linear equations, we can obtain the resistance of multilayer CrPS$_4$, 
\begin{equation}
R=(V_1-V_n)/I=V_1/I.
\end{equation}

\comment{Here we focus on the case of a 6-layer CrPS$_4$ sample. By setting ${a_1}=0.1,{a_2}=2, \sigma=0.1 \Delta E, d/\lambda_z=1$, we obtain the MR as shown in Supplementary Fig. 17. As the magnetic field increases, multiple resistance peaks (denoted by P1, P2, and P3) appear. These resistance peaks result from a competition between wavefunction contraction, which increases resistance, and an increase in the density of defect states, which decreases resistance, under the applied magnetic field.}

\begin{figure}[H]
\addtocounter{figure}{13}
\centering
\includegraphics[width =0.5\linewidth]{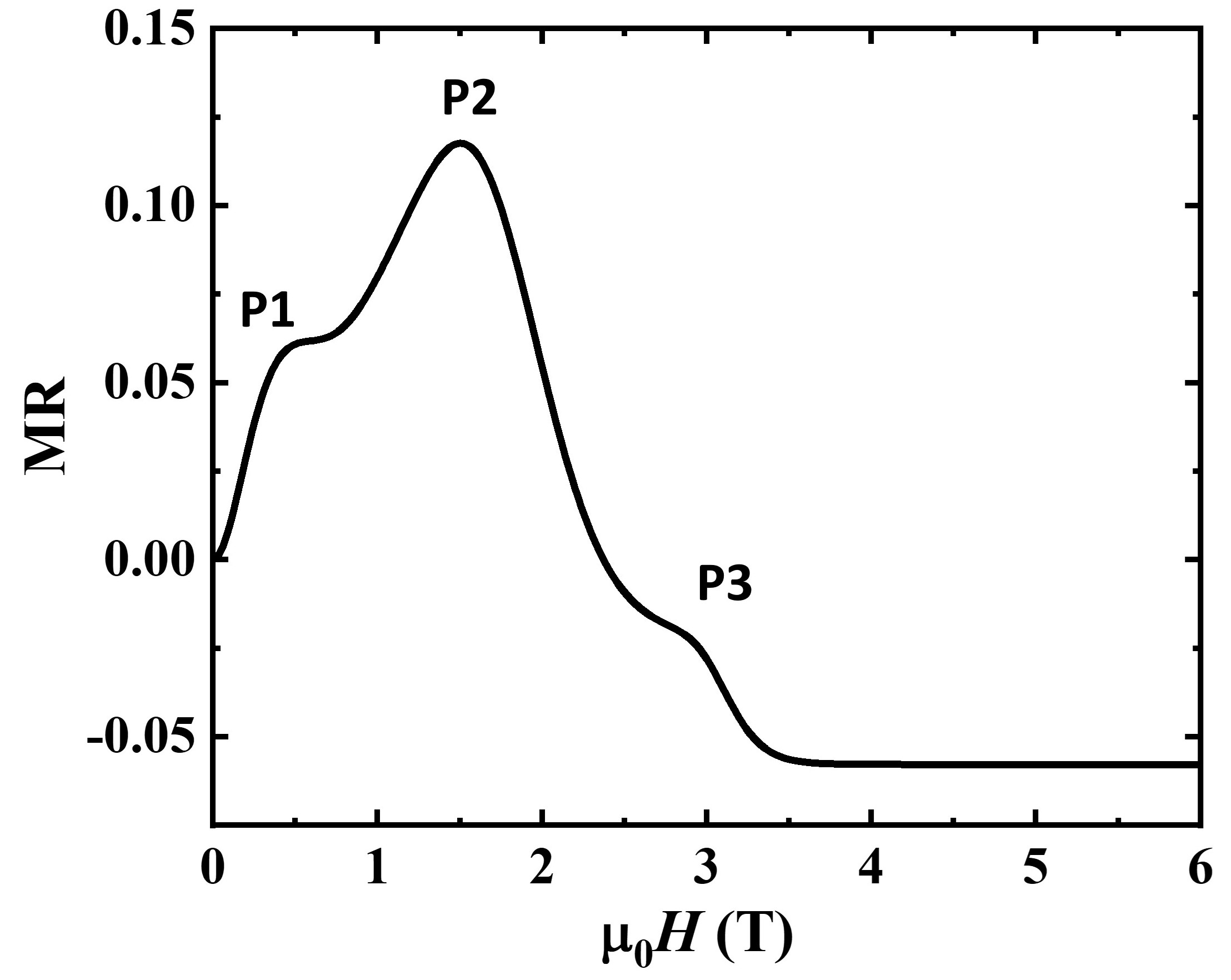}
\caption{Simulated MR for six-layer CrPS4 system with three peaks labeled by P1, P2, P3.} 
\end{figure}

The next step is to investigate how bias voltage affects the peak locations. The energy difference between different layers can be tuned by the vertical electric field $V$ ~\cite{VRHEfield0,VRHEfield2,VRHEfield3}, we assume a uniform distribution of electric field along $z$ direction,
\begin{equation}
\Delta {E_{ij}} = {\sin ^2}{\theta  \over 2}\Delta E - {{j - i} \over n}eV .
\end{equation}
The DOS function $f(H)$ is therefore modified as following,
\begin{equation}
f(H) = \exp \left({{{\Delta E^2} \over {2\sigma^2}} \left({1-{\rm Max} \left(0,{\sin ^2}{\theta  \over 2} -{{j - i} \over n}V'\right)^2 }\right)}\right).
\end{equation}
Here the function Max takes the maximum value of 0 and  ${\sin ^2}{\theta  \over 2} -{{j - i} \over n}V'$, where $V'=eV/\Delta E$. The existence of electric field $V'$ would increase the DOS and therefore the dimensionless hopping range $\xi$ would decrease. The negative correlation between electric field $V$ and $\xi$ is well known, it is reported $\xi \propto 1/V^\alpha$ with $\alpha$ varies from 1/2, 1 and 2 as in reference~\cite{VRHEfield0,VRHEfield2,VRHEfield3}. Here we focus on how the MR peaks shift under various electric field $V'$, MR curves are summarized in Supplementary Fig. 18. \comment{As the electric field increases, MR oscillations become less pronounced, and the peak positions shift to lower magnetic fields, as shown in the inset of Supplementary Fig. 18. Under a high electric field, such as $V' > 0.1$, the P3 peak disappears, and the overall positive MR transitions to a negative MR. These simulation results qualitatively agree well with the experimental findings.}

\begin{figure}[H]
\centering
\includegraphics[width =0.6\linewidth]{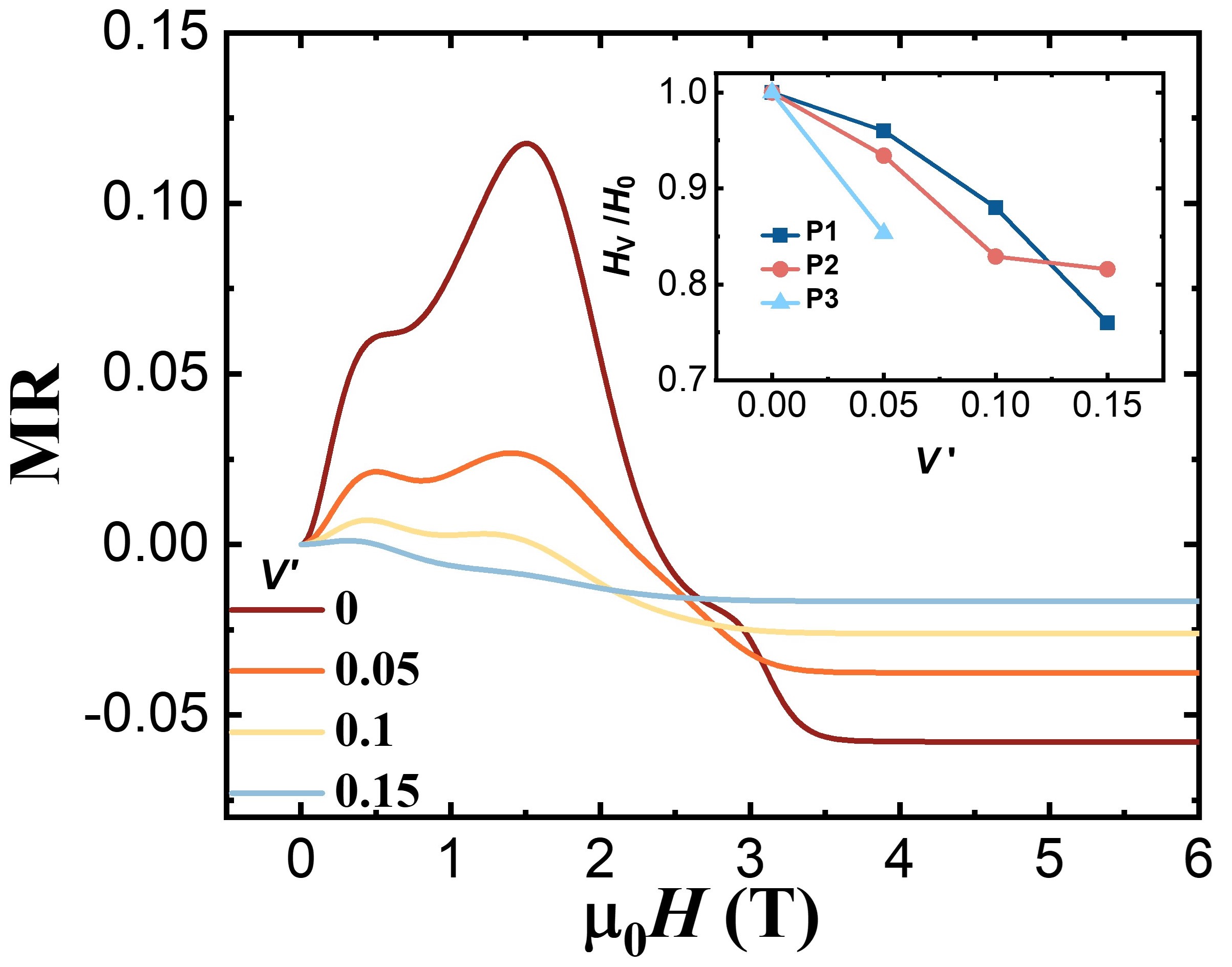}
\caption{Simulated MR for six-layer CrPS$_4$ system under various electric fields. Inset summarize the peak position shifts under electric fields.} 
\end{figure}

\comment{By accounting for wavefunction contraction and DOS variation effects under a magnetic field, we successfully reproduce MR oscillations and electric field modulations that qualitatively align with our experimental observations. However, we are unable to obtain peaks at positions matching those in the experimental results, indicating that further refinement is needed for quantitative agreement.} To enhance the interpretation, we propose that an interference term  arises in spin-polarized hopping conduction systems. This interference term, denoted by $\Psi(\theta)$, allowing Eq. (20) to be rewritten as:

\begin{equation}
G_{ij} = \exp[-(\xi_{ij}+|i-j|d/\lambda_z)] \cdot \Psi(\theta),
\end{equation}

The introduction of $\Psi(\theta)$ is inspired by the theoretical work in Ref. \onlinecite{spinphase}, where it is proved that the spin Berry phase emerges during any cyclic quantum evolution. The phase is given by $\pi(1 - \cos\theta)$, with $\theta$ defined as the angle between the spin direction and the external magnetic field. While the exact reason why Spin Berry Phase show up in CrPS$_4$ needs further investigations, we demonstrate that our model yields MR curves that closely match the experimental results after incorporating it. Since both spin-up and spin-down bands are involved, the spin follows different evolutionary paths with opposite spin directions during hopping, and the resulting spin Berry phase difference $\phi_{\theta}$ is given by:
\begin{equation}
\phi_{\theta} = \sum_{k=i+1}^{j}\pi\left(1-\cos \theta_k\right),
\end{equation}
where $\theta_k$ denotes angle between external magnetic field and magnetization of $k$th layer and can be obtained from linear chain model as describe in Supplementary Note 4. Given that our experimental data exhibit up to three peaks for thicker samples, such as the 18-layer CrPS$_4$, we infer that the coherence length associated with the spin Berry phase is limited to approximately 6 layers. Therefore, we only consider maximum hopping distance of 6 layers, i.e. $|i-j| \le 6$, where $i$ and $j$ label the layer indices. Within the coherence length, the phase factor $\Psi(\theta)$ is related to the Berry phase difference $\phi_{\theta}$ by,
\begin{equation}
\Phi_{\theta} = \left|P+(1-P)\exp(i\phi_{\theta})  \right|.
\end{equation}
Here $P$ denotes the spin polarization of defect states.

With the spin Berry phase-induced interference term, we show that the peak positions align closely with the experimental results. By setting  ${a_1}=0.2,{a_2}=1, \sigma=0.25 \Delta E, d/\lambda_z=0.7, P=0.75$, we obtain the MR curve shown in Supplementary Fig. 19(a). Two pairs of peaks near 3 T and 6 T are observed with locations closely matching experimental results. Additionally, we  simulated MR curves for thicker CrPS$_4$ samples, specifically the 10- and 12-layer systems. Since the measured 10- and 12-layer samples exhibit higher conductivity than the 6-layer sample, we adjusted the localization length by setting $d/\lambda_z = 0.23$ while keeping the other parameters unchanged. The resulting MR curves display three peaks, all of which are in good agreement with the experimental data.

\begin{figure}[H]
\centering
\includegraphics[width =1\linewidth]{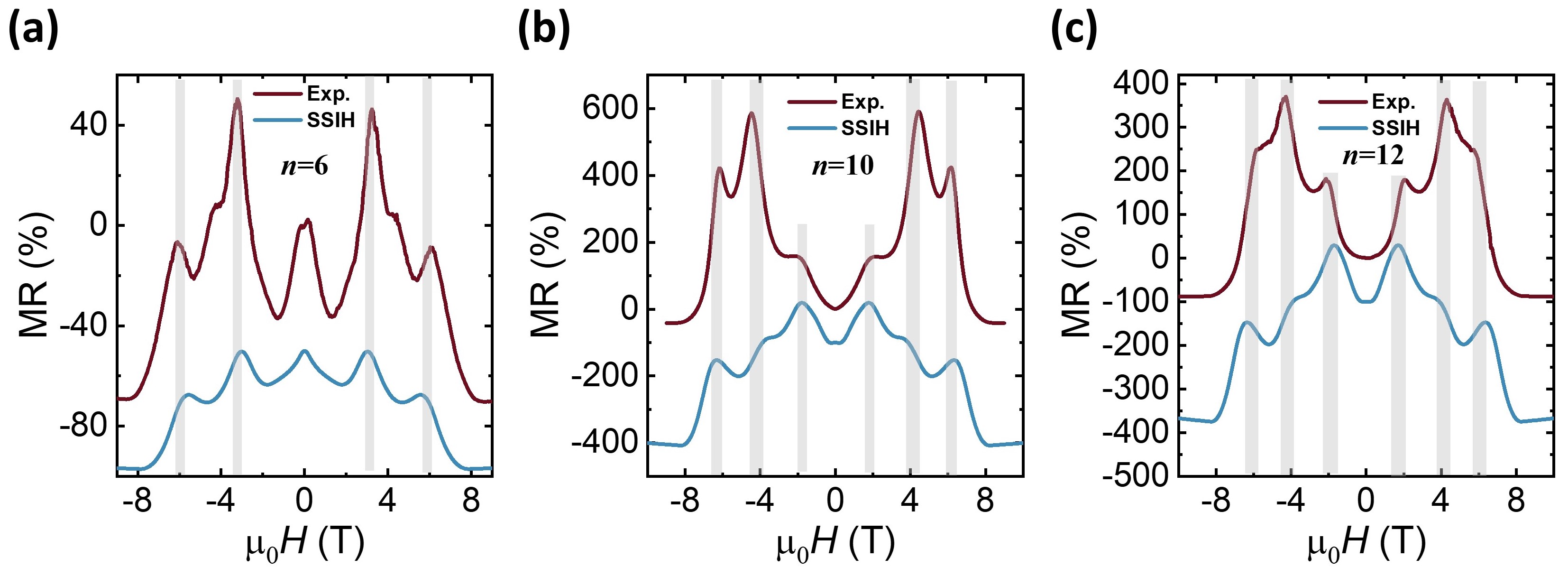}
\caption{\textbf Simulated MR curves for the (a) 6-layer, (b) 10-layer and (c) 12-layer CrPS$_4$ system are represented by the blue curves. For clarity, the MR curves are shifted vertically, and the MR curves for the 10- and 12-layer systems, shown in (b) and (c), are magnified by a factor of 10. The red curves denote the corresponding experimental results for comparison.} 
\end{figure}

In addition, we can also investigate the bias voltage effect. With the increase of electric field, the peak position P2 and P3 are shifted to lower magnetic field as shown by the inset of Supplementary Fig. 20. These simulation results qualitatively agrees well with the experimental results.

\begin{figure}[H]
\centering
\includegraphics[width =0.6\linewidth]{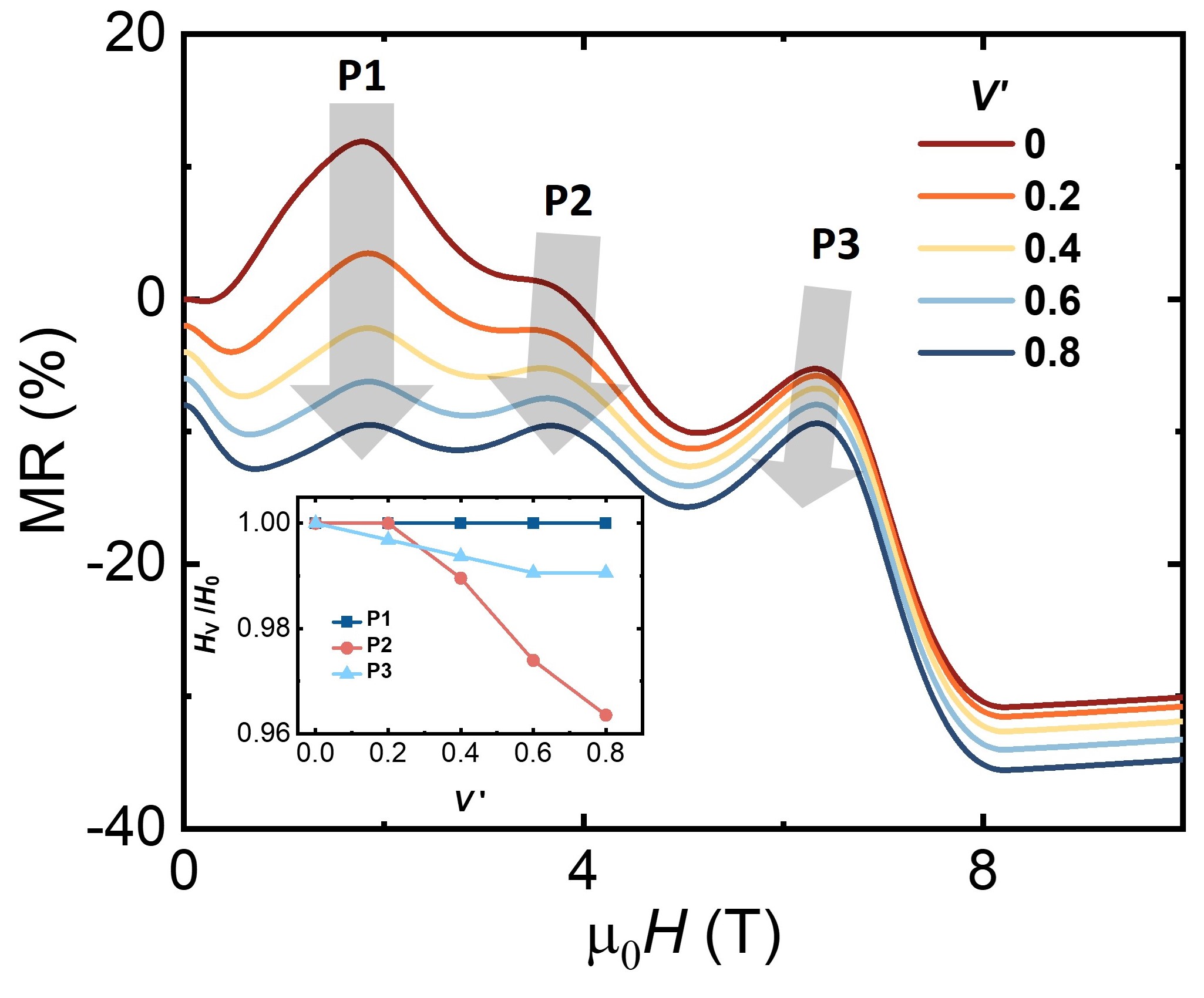}
\caption{ Simulated MR for 10-layer CrPS$_4$ system under various electric fields. Inset summarize the peak position shifts under electric fields.} 
\end{figure}

\comment{In summary, the competing effects of wavefunction contraction and DOS variation under a magnetic field can lead to MR oscillations, which are effectively modulated by the electric field. This model qualitatively aligns with our experimental observations; however, it does not accurately predict the exact peak positions or MR magnitude. Two possible issues could explain this discrepancy:(a) accurately predicting peak positions and magnitudes requires a more detailed evaluation of how the magnetic field influences the DOS and localization length of defect states in 2+1 dimensions, and (b) within a single hopping process, the spin Berry phase might also affect the hopping probability, as discussed above. The mechanism governing variations in the spin Berry phase in CrPS$_4$, which is absent in other A-type 2D antiferromagnets such as CrI$_3$ and CrCl$_3$, remains unclear. Future research, including defect engineering in CrPS$_4$, is necessary to provide more definitive insights.}

\clearpage

\clearpage
\newpage
\bibliographystyle{mynaturemag}
\bibliography{biblio}